# Quantum Sensing with Triplet Pair States: A Theoretical Study

Maria Grazia Concilio[1]*, Yiwen Wang[1], Siyuan Wang[1], Xueqian Kong[1]*

[1]*Institute of Translational Medicine, Shanghai Jiao Tong University, 200240 Shanghai, China*

## Abstract

Molecular quantum sensors represent a promising frontier for the detection of nuclear magnetic resonance (NMR) signals and alternating current (AC) magnetic fields at the nanoscale, potentially reaching single-proton sensitivity. Although the triplet states $(\hat{T})$ of molecular pentacene provide a viable sensing architecture, the triplet pair states $(\hat{T}\hat{T})$ produced by singlet fission of pentacene dimers could enable more flexible quantum manipulations through entanglement. In this work, we model the quantum sensing efficacy of a spin-polarized quintet manifold ${}^{5}(\hat{T}\hat{T})_{0,\pm1}$in a photoexcited pentacene dimer generated via intramolecular singlet fission (SF). Using a Lindblad master equation approach, we simulate the evolution of the triplet-pair state under standard dynamical decoupling (DD) sequences—including spin echo (SE), XY4, and XY8 and provide a direct performance comparison to the traditional pentacene monomer benchmark. While both architectures exhibit comparable sensitivity for isolated single-spin detection, our findings indicate that the dimer architecture provides a superior interaction cross-section for detecting small ensembles of nuclear spins. Analytical expressions derived for fluorescence modulation demonstrate that sensitivity is optimized in the low-magnetic field regime (≤0.01 T) and scales with the number of pulses in the sensing protocol. This study establishes a theoretical baseline for utilizing high-spin multi-excitonic states as chemically tunable, high-sensitivity quantum probes.

## Introduction:

Quantum sensors represent a paradigm shift in metrology, offering unprecedented sensitivity and spatial resolution for measuring magnetic, electric, and gravitational fields, as well as local physical parameters such as pressure, temperature, and pH. [1, 2] While traditional quantum sensing has been dominated by solid-state defect centers—most notably the nitrogen-vacancy (NV) center in diamond [2-6]—there is increasing interest in molecular spin systems due to their chemical tunability and the ability to fabricate atomically defined sensor assemblies at scale. [3, 7-9] Photoexcited organic chromophores, such as pentacene, have emerged as a significant platform for this purpose, offering optical initialization and readout at room temperature through state-dependent fluorescence contrast. [10-12]

Recent investigations have successfully utilized pentacene monomers to detect single $^{1}H$ and $^{13}C$ nuclei using optically detected magnetic resonance (ODMR) and dynamical decoupling (DD) pulse sequences. [4, 10, 12, 13] However, a key strategy to enhance sensitivity involves the exploitation of entanglement between triplet-pair states in covalent dimers (the specific example investigated in this study is shown in Figure 1A). [14] These systems undergo intramolecular singlet fission (SF), [15, 16] a spin-allowed process where a photogenerated singlet exciton $\hat{S}_1\hat{S}_0$ is converted into a correlated triplet pair (Figure 1B). In the typical four-stage SF model, the initially formed singlet-character triplet pair $^1(\hat{T}\hat{T})_0$ evolves into a high-spin quintet manifold $^5(\hat{T}\hat{T})_{0,\pm1,\pm2}$.[17, 18] In the last stage of the SF, the triplet pair dissociates into two isolated triplets $\hat{T}_1 + \hat{T}_1$ which later decay away through a recombination process. The transient-state $^5(\hat{T}\hat{T})_0$ state is particularly attractive for quantum information science as it provides a multi-level spin system (qubit) that is highly entangled and addressable via microwave pulses.[19-21]

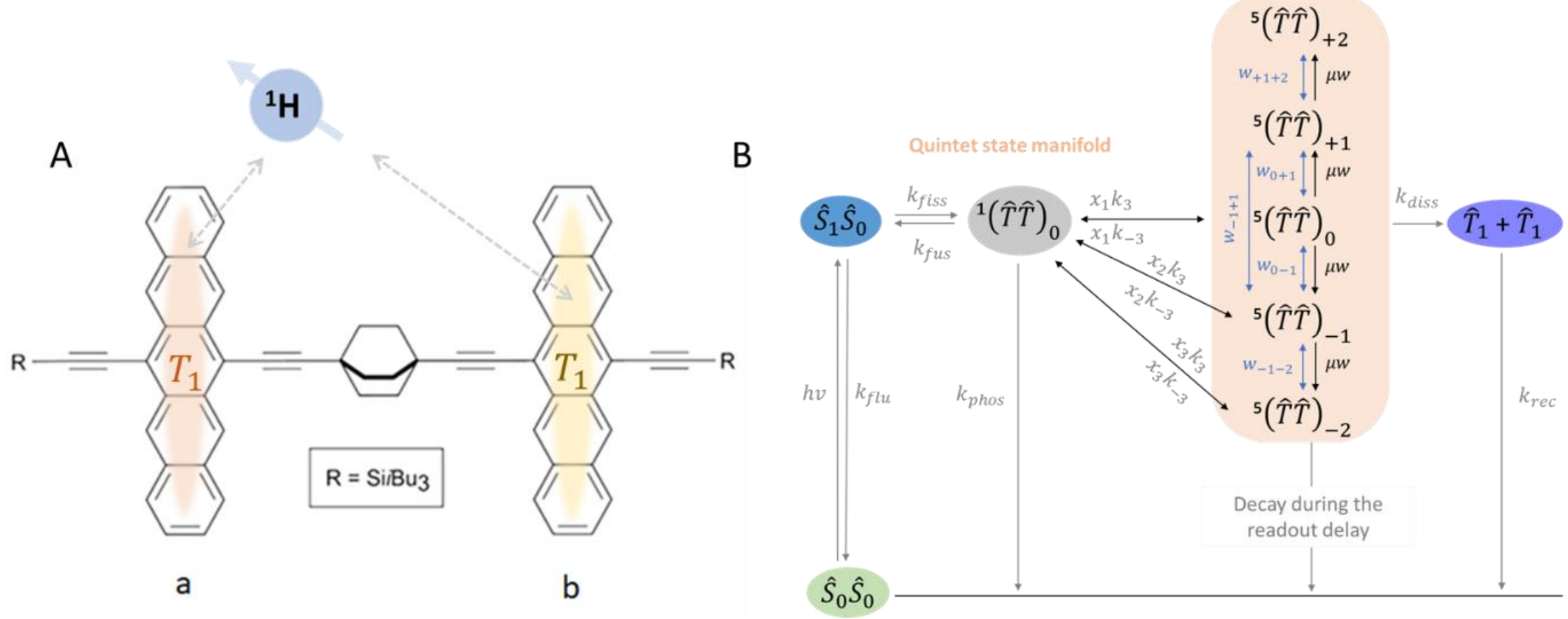

**Figure 1:** A) Chemical structure of the pentacene dimer composed of two monomers a and b. $T_1$ represents the triplet state composed by two electrons delocalized in each pentacene ring. The triplets interact with a proton through dipolar interaction. B) The SF scheme, where the constants $k_{hv}$, $k_{fis}$, $k_{fus}$, $k_{diss}$, $k_{phos}$ and $k_{rec}$ represent the rates of the laser excitation, fission, fusion, dissociation, phosphorescence and recombination respectively, $k_3$ and $k_{-3}$ represent the rate from the state $^1(\hat{T}\hat{T})_0$ to the quintet subspace and from the quintet subspace the state $^1(\hat{T}\hat{T})_0$ respectively. μw represents the microwave irradiation. $w_{0\pm1}$, $w_{+1-1}$ and $w_{\pm1\pm2}$ represent the longitudinal relaxation rates between the triplet sublevels. $x_1$, $x_2$ and $x_3$ correspond to the populations of $^5(\hat{T}\hat{T})_0$, $^5(\hat{T}\hat{T})_{-1}$and $^5(\hat{T}\hat{T})_{-2}$, respectively.

Beyond photovoltaics, where SF acts as an exciton multiplier [15, 16, 22], the unique spin degrees of freedom in pentacene dimers are being explored for quantum technology [17, 23] and enhanced NMR sensitivity via triplet-J-driven dynamic nuclear polarization (JDNP). [14] Despite this potential, a comprehensive theoretical framework that bridges the gap between SF kinetics and coherent sensing protocols remains underdeveloped.

In this work, we utilize the Lindblad master equation to model the interplay between the incoherent kinetics of SF and the coherent spin dynamics of the triplet-pair manifold in the presence of nuclear spin environments. We focus on the coherent control of the $^5(\hat{T}\hat{T})_0$ to $^5(\hat{T}\hat{T})_{+1}$transition and evaluate the efficacy of standard sensing protocols based on DD sequences—SE, XY4, and XY8—for the detection of single nuclear spins, nuclear ensembles (in Figure 2),[3] and alternating current (AC) magnetic fields. [24] By providing a comparative analysis of the dimer and monomer architectures, we identify the critical regimes where the high-spin character of the triplet pair provides a definitive sensing advantage.

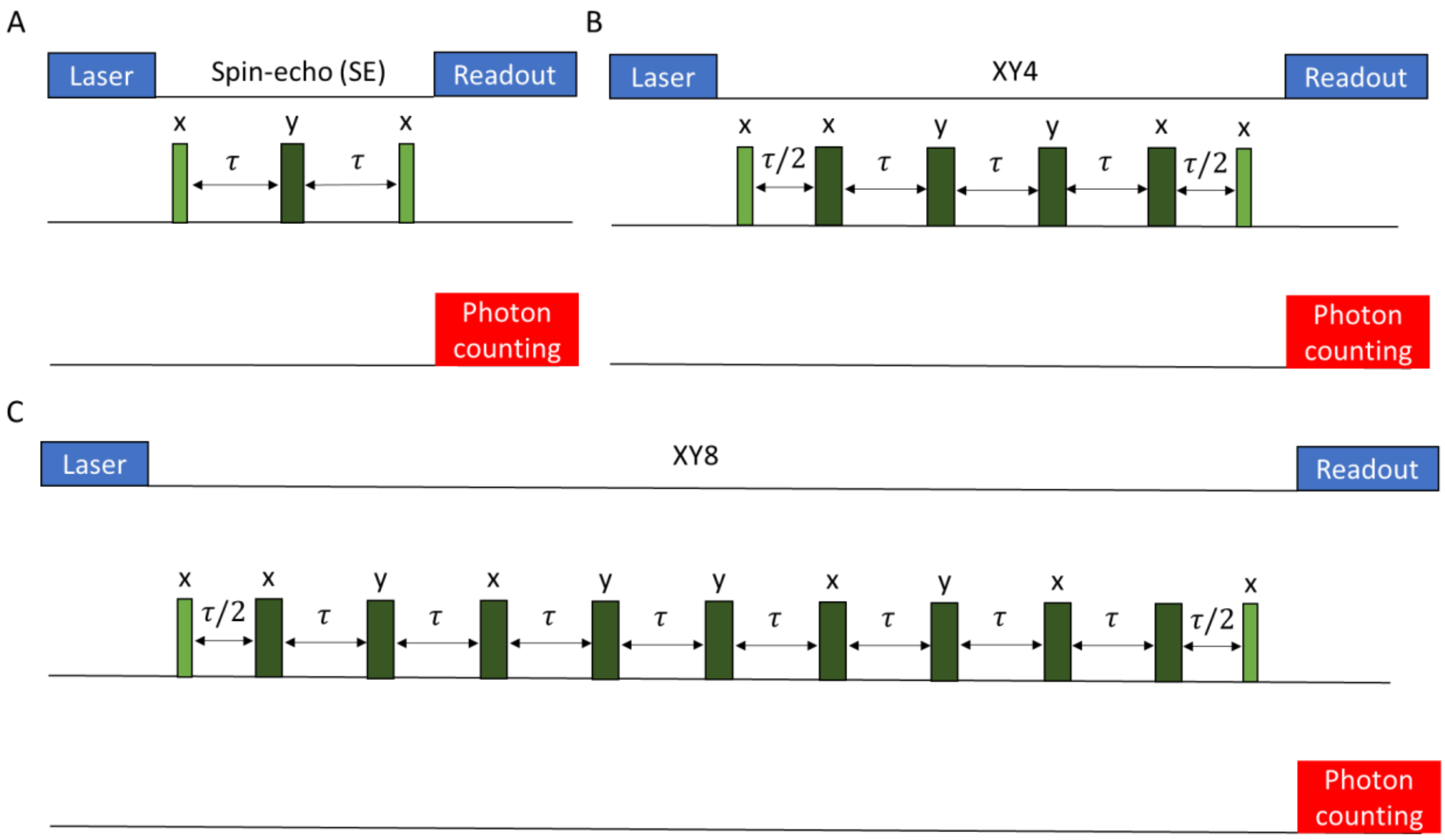

**Figure 2:** (A) Spin-echo (SE), (B) XY4 and (C) XY8 pulse sequences. The light and dark green boxes represent the π/2 and π pulses respectively, τ is the inter-pulse delay.

**Spin systems and methodology:**

A pentacene dimer with structure shown in Figure 1A was considered as the model systems in this work. The SF process that generates polarized triplet pair states, shown in Figure 1B, was simulated using the Lindblad master equation, used previously to model the spin dynamics of NV-centers and pentacene dimers, [25] Eq. 1:

$$\frac{\partial \vec{\rho}}{\partial t} = -i[H, \vec{\rho}]_{-} + \sum_k \left( \mathcal{L}_k \rho \mathcal{L}_k^{\dagger} - \frac{1}{2} \left[ \mathcal{L}_k^{\dagger} \mathcal{L}_k, \rho \right]_{+} \right) \tag{1}$$

where ρ is the density matrix, H is the spin Hamiltonian used to describe the coherent dynamics arising from couplings present in the system and $\mathcal{L}_k$ are the collapse operators used to describe the incoherent dynamics arising from kinetics and relaxation processes, $[\cdot,\cdot]_{-}$ and $[\cdot,\cdot]_{+}$ are the commutators and anticommutators respectively. The details about the simulation methodology are given in the section A in the supporting information (SI).

In the single crystal simulation, the pentacene dimers are assumed to stack in the parallel fashion with π-π interactions between neighboring molecules. [26, 27] So, the ZFS- and the g-tensors are parallel to each other and are aligned with the molecular frame. We also assume that the spin centers align with the crystal axes, and the crystal is oriented parallel to the magnetic field in the spectrometer.

We also implement a powder simulation in Spinach package [28], using a 1 angle repulsion powder grid with 400 points required for the powder averaging.

The electron $g$-tensors are set with the eigenvalues taken from those found in pentacene in a single crystal of naphthalene. [29, 30] The zero-field splitting (ZFS) parameters D and E are set equal to 1139 MHz and 60 MHz as those observed in pentacene dimers. [21, 31] The inter-triplet exchange coupling $J_{ex}$ of acetylene / phenylene linked pentacene was observed to be around 15 GHz. [31]

The kinetics constants involved in the SF model shown in Figure 1B were determined experimentally using transient absorption (TA) spectroscopy and the time resolved electron paramagnetic resonance (TREPR) (see ref. [15, 16]). The spin lattice relaxation at room temperature (RT) between the triplet sublevels in pentacene monomers were determined experimentally: $\hat{T}_0 \leftrightarrow \hat{T}_{+1}$ = 1.1×10$^4$ Hz, $\hat{T}_0 \leftrightarrow \hat{T}_{-1}$ = 2.2×10$^4$ Hz and $\hat{T}_{-1} \leftrightarrow \hat{T}_{+1}$ = 0.4×10$^4$ Hz. [10, 13] In this work we focus on single transitions $^5(\hat{T}\hat{T})_0 \leftrightarrow {}^5(\hat{T}\hat{T})_{\pm1}$(i.e. $w_{0,\pm1}$), and $^5(\hat{T}\hat{T})_{-1} \leftrightarrow {}^5(\hat{T}\hat{T})_{-2}$ (i.e. $w_{-1,-2}$) and assume that relaxation rates between these states are comparable to those observed in the pentacene monomer. The rates $w_{0,\pm1}$, $w_{-1,+1}$ and $w_{\pm1,\pm2}$ are set to 1×10$^4$ Hz. The presence of asymmetric rates between the other states [21] would have a minor effect on the populations of the states considered in this work. The decoherence time, $T_2$, is set between 1 μs – 10 μs for the pentacene (see ref. [10, 12]) and 0.1 μs - 10 μs for the pentacene dimer (see ref. [32]), respectively.

The populations of the quintet states $^5(\hat{T}\hat{T})_0$, $^5(\hat{T}\hat{T})_{-1}$and $^5(\hat{T}\hat{T})_{-2}$ equal to 0.60, 0.10 and 0.30 respectively, which are comparable to those used in other pentacene dimers. [20] With these numbers, the TREPR spectrum of the pentacene dimer (shown in ref. [16]) can be simulated using a single configuration as described above. The relevant simulation parameters of the pentacene dimer considered in this work are summarized in the Table 1:

**Table 1:** Parameters for the triplet-pair / proton spin system used in the simulations.

| **Parameter** | **Value** |
|---|---|
| $^1$H chemical shift tensor, ppm | [5 5 5] |
| Triplet a $g$-tensor eigenvalues, [xx yy zz] / Bohr magneton | [2.0015 2.0009 2.0005] |
| Triplet a $g$-tensor, ZYZ active Euler angles / rad | [0.0 0.0 0.0] |
| Triplet b $g$-tensor eigenvalues, [xx yy zz] / Bohr magneton | [2.0015 2.0009 2.0005] |
| Triplet b $g$-tensor, ZYZ active Euler angles / rad | [0.0 0.0 0.0] |
| $^1$H coordinates [x y z] / Å* | [2.2 0.0 9] |
| Triplet 1 and triplet 2 coordinates, [x y z] / Å* | [0 0 -7] and [0 0 7] |
| Zero field splitting (ZFS) parameters, D and E / MHz | 1139 and 60 |
| Inter-triplet exchange coupling $J_{ex}$ / GHz | 15 |

| Inter-triplet dipolar interaction $d$ /MHz | 23 |
|---|---|
| Kinetics constants:<br>$k_{\text{flu}}$, $k_{\text{fis}}$, $k_{\text{fus}}$,<br>$k_{\text{phos}}$, $k_3$, $k_{-3}$,<br>$k_{\text{diss}}$, $k_{\text{rec}}$ and $k_{\text{hv}}$, / $s^{-1}$ | $5.1\times10^7$, $3.3\times10^8$, $1\times10^6$,<br>$4.1\times10^6$, $6.7\times10^6$, $1.8\times10^6$,<br>$6.7\times10^5$, $2.5\times10^4$, $3.1\times10^7$ |
| Magnetic field, $B_0$ / T | From 0.01 to 1 |

[1*] Spin interactions were computed from the Cartesian coordinates and the spins interaction matrices, following the irreducible spherical tensor notation, as done also in the Spinach software. [28]

Simulations of the pentacene monomer are performed by removing a triplet and inter-triplet coupling parameters. We did not take into account of the correlated singlet–triplet pair states, since they were not observed in the TREPR spectrum of the molecule under consideration. [15, 16] On the other hand, singlet–triplet pair states are expected to arise from the states $^3(\hat{T}\hat{T})_{0,\pm1}$, in molecules where heavy atoms are present. [33] Simulations of ODMR experiments are carried out at 1 T, while simulations of DD experiments are carried out between 0.01 T and 1T. The simulation scripts are written in both MATLAB and Python and can be provide upon request.

## Results and discussion:

### 1) The simulations of ODMR spectra:

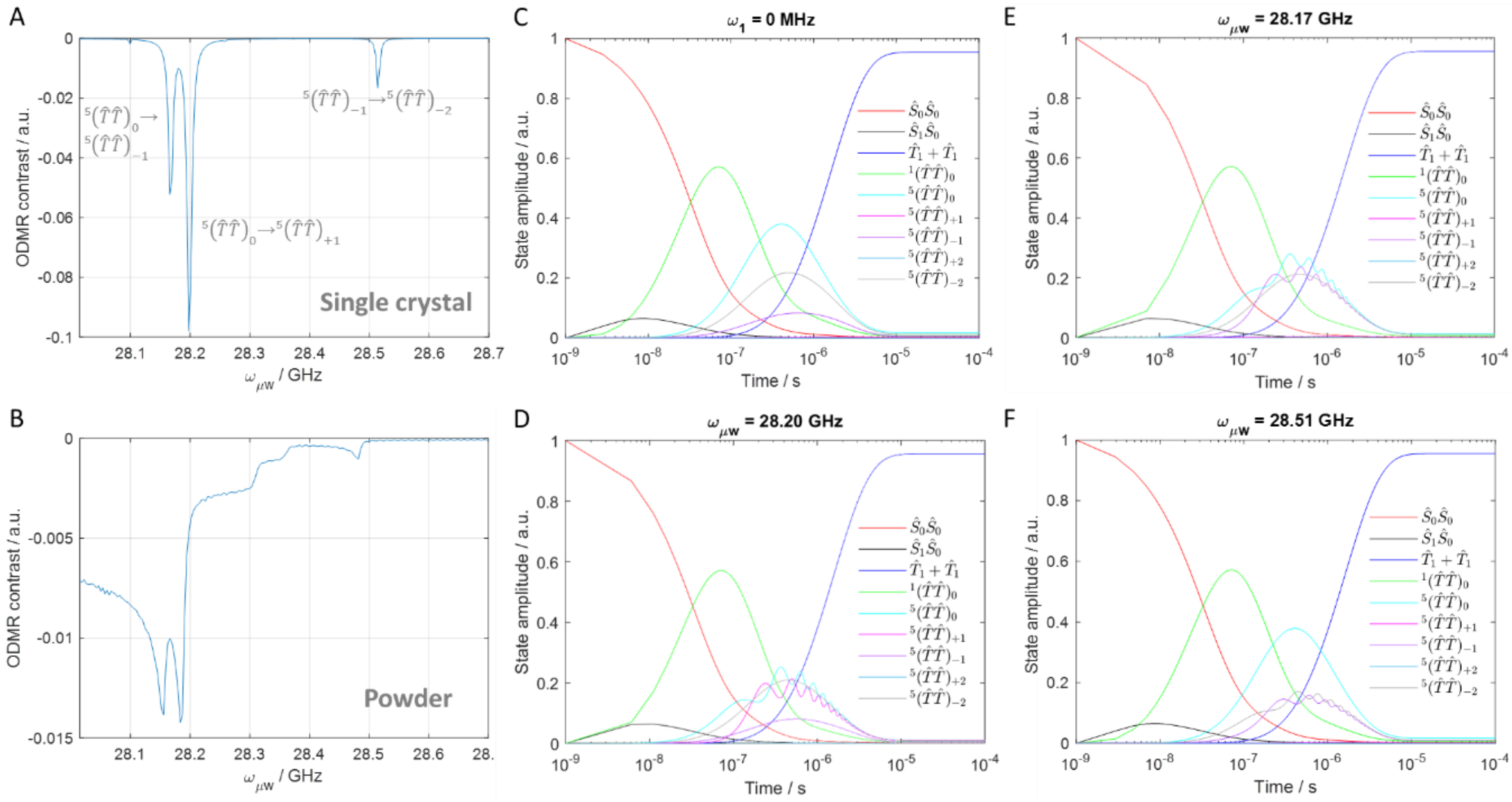

**Figure 3:** ODMR contrast as a function of the microwave frequency $\omega_{\mu w}$ at 1 T, using a single crystal (A) and a powder (B). C) Time – domain simulations showing the dynamics of the states involved in the SF setting the microwave power $\omega_1$ equal to zero. D) Time – domain simulations showing the dynamics of the states involved in the SF setting $\omega_1$ and $\omega_{\mu w}$ equal to 1 MHz and 28.20 GHz, corresponding to the transition $^5(\hat{T}\hat{T})_0\rightarrow{}^5(\hat{T}\hat{T})_{+1}$. E) Time – domain simulations showing the dynamics

of the states involved in the SF setting $\omega_1$ and $\omega_{\mu w}$ equal to 1 MHz and 28.17 GHz, corresponding to the transition ${}^5(\hat{T}\hat{T})_0 \rightarrow {}^5(\hat{T}\hat{T})_{-1}$. F) Time – domain simulations showing the dynamics of the states involved in the SF setting $\omega_1$ and $\omega_{\mu w}$ equal to 1 MHz and 28.51 GHz, corresponding to the transition ${}^5(\hat{T}\hat{T})_{-1} \rightarrow {}^5(\hat{T}\hat{T})_{-2}$. The time dependent Hamiltonian is shown in the section A in the SI. The simulation parameters are given in the Table 1. In the simulation, the distances between the proton and the two pentacene centers are 3 Å and 16 Å, respectively.

Simulations were performed using a pseudo-secular Hamiltonian, in which a rotating frame approximation is applied to the triplet states. In this approximation, the Zeeman carrier frequency is subtracted from the trace of the Zeeman interaction tensor. The ODMR contrast was calculated from the amplitude of the $\hat{S}_1\hat{S}_0$ state at the steady state resulting from the overall photocycle in Figure 1B, in according to:

$$\text{ODMR} = \frac{\hat{S}_1\hat{S}_{0,\mu\text{w on}} - \hat{S}_1\hat{S}_{0,\mu\text{w off}}}{\hat{S}_1\hat{S}_{0,\mu\text{w off}}} \tag{2}$$

where $\hat{S}_1\hat{S}_{0,\mu\text{w on}}$ and $\hat{S}_1\hat{S}_{0,\mu\text{w off}}$ correspond to the population of $\hat{S}_1\hat{S}_0$ obtained when the μw are on and off respectively. The position of the transitions is mostly determined by the magnitude of $\omega_E$, $J_{\text{ex}}$ and D, and can be determined using Eq. 3 below. The ZFS parameter E has a minor effect on the ODMR contract and to simplify the analysis we omitted it. The energies of the eigenstates, obtained through diagonalization of the Hamiltonian in Eq. S6, were given in the SI. The allowed transition, that fulfill $\Delta m_s = \pm 1$, correspond to ${}^3(\hat{T}\hat{T})_0 \rightarrow {}^3(\hat{T}\hat{T})_{\pm 1}$, ${}^5(\hat{T}\hat{T})_0 \rightarrow {}^5(\hat{T}\hat{T})_{\pm 1}$ and ${}^5(\hat{T}\hat{T})_{\pm 1} \rightarrow {}^5(\hat{T}\hat{T})_{\pm 2}$. In this work, we focus on the transitions ${}^5(\hat{T}\hat{T})_0 \rightarrow {}^5(\hat{T}\hat{T})_{\pm 1}$ that occur at the transition frequencies:

$$\Delta E_{\mathbf{5\pm 10}} = \frac{3J_{\text{ex}}}{2} - \frac{1}{2}\sqrt{4(3\text{d}^2 - 2\text{dD} + \text{D}^2) - 4(3\text{d} + \text{D})J_{\text{ex}} + 9J_{\text{ex}}^2} \pm \omega_E \text{ for } {}^5(\hat{T}\hat{T})_0 \rightarrow {}^5(\hat{T}\hat{T})_{\pm 1} \tag{3}$$

where d is the inter-triplet dipolar coupling corresponding to the secular part of the anisotropic coupling between the two triplet states, and $\omega_E$ is the electron Larmor frequency. Eq. 3 shows the possibility of obtaining information on the magnitude of the $J_{\text{ex}}$ for the position of the transitions ${}^5(\hat{T}\hat{T})_0 \rightarrow {}^5(\hat{T}\hat{T})_{\pm 1}$, whose difference is equal to $|2\omega_E|$, as observed also in the case of the pentacene monomer between the transitions $\hat{T}_0 \rightarrow \hat{T}_{\pm 1}$. [34]

The ODMR of the pentacene monomer was already discussed in ref. [10, 13] therefore we focus only on the pentacene dimer. We simulate the ODMR spectra for both single crystal (Figure 3A) and powder (Figure 3B) of the pentacene dimer at the magnetic field of 1 T. The ODMR spectrum of the single crystal is composed of three dips in the PL at 28.17, 28.20 and 28.51 GHz, corresponding to the transitions ${}^5(\hat{T}\hat{T})_0 \rightarrow {}^5(\hat{T}\hat{T})_{-1}$, ${}^5(\hat{T}\hat{T})_0 \rightarrow {}^5(\hat{T}\hat{T})_{+1}$ and ${}^5(\hat{T}\hat{T})_{-1} \rightarrow {}^5(\hat{T}\hat{T})_{-2}$ respectively. For sake of clarity, we add the value of the Zeeman carrier frequency at 1 T (equal to 28 GHz) to the x-axis. The ODMR powder spectrum is broader due to the large number of configurations contributing to the signal and the two dominant dips shift slightly towards lower frequencies. For the powder sample, transition frequencies ${}^5(\hat{T}\hat{T})_0 \rightarrow {}^5(\hat{T}\hat{T})_{\pm 1}$ and ${}^5(\hat{T}\hat{T})_{-1} \rightarrow {}^5(\hat{T}\hat{T})_{-2}$ change slightly with the α and γ angle of the powder grid (the Euler angles with respect to $B_0$ field), and massively with the β angle (see Figure S1 in the supporting information for more details). Due to the computational cost of the powder sample, the later analyses are focused only on the single crystal sample.

Figures 3C-3F show the dynamics of the individual states involved in the SF process, in the absence (Figure 3C) and in the presence of μw irradiation at the three transition frequencies 28.17 GHz (Figure 3D), 28.20 GHz (Figure 3E) and 28.51 GHz (Figure 3F). By setting the initial conditions equal to the state $\hat{S}_0\hat{S}_0$, the states $\hat{S}_1\hat{S}_0$, ${}^5(\hat{T}\hat{T})_{0,-1,-2}$ and $\hat{T}_1 + \hat{T}_1$ are created sequentially in according the kinetics model shown in Figure 1 and Eq. S5. The state population is dominated by the SF kinetics rather than the longitudinal relaxation ranging between 10 – 100 μs. The state ${}^5(\hat{T}\hat{T})_0$ reaches its maximum amplitude about 0.4 μs after the laser irradiation, and a steady state population after 10 μs (see Figure 3C). In the presence of an irradiation at the transition frequencies, it is possible to observe Rabi oscillations between the transitions ${}^5(\hat{T}\hat{T})_0 \rightarrow {}^5(\hat{T}\hat{T})_{\pm 1}$ and ${}^5(\hat{T}\hat{T})_{-1} \rightarrow {}^5(\hat{T}\hat{T})_{-2}$ in Figures 3D-F, demonstrating the possibility of realizing coherent spin control in these systems.

### 2) The applications of dynamic decoupling sequences:

Numerical simulations show that both the pentacene and the pentacene dimer can be used to detect a single nuclear spin using the SE, XY4 and XY8 sequences. The presence of nuclear spin leads to dips in the fluorescence, corresponding to the state $\hat{T}_0$ and to the state ${}^5(\hat{T}\hat{T})_0$ in the case of the pentacene monomer and dimer respectively, as a function of the τ delay, see Figure 4.

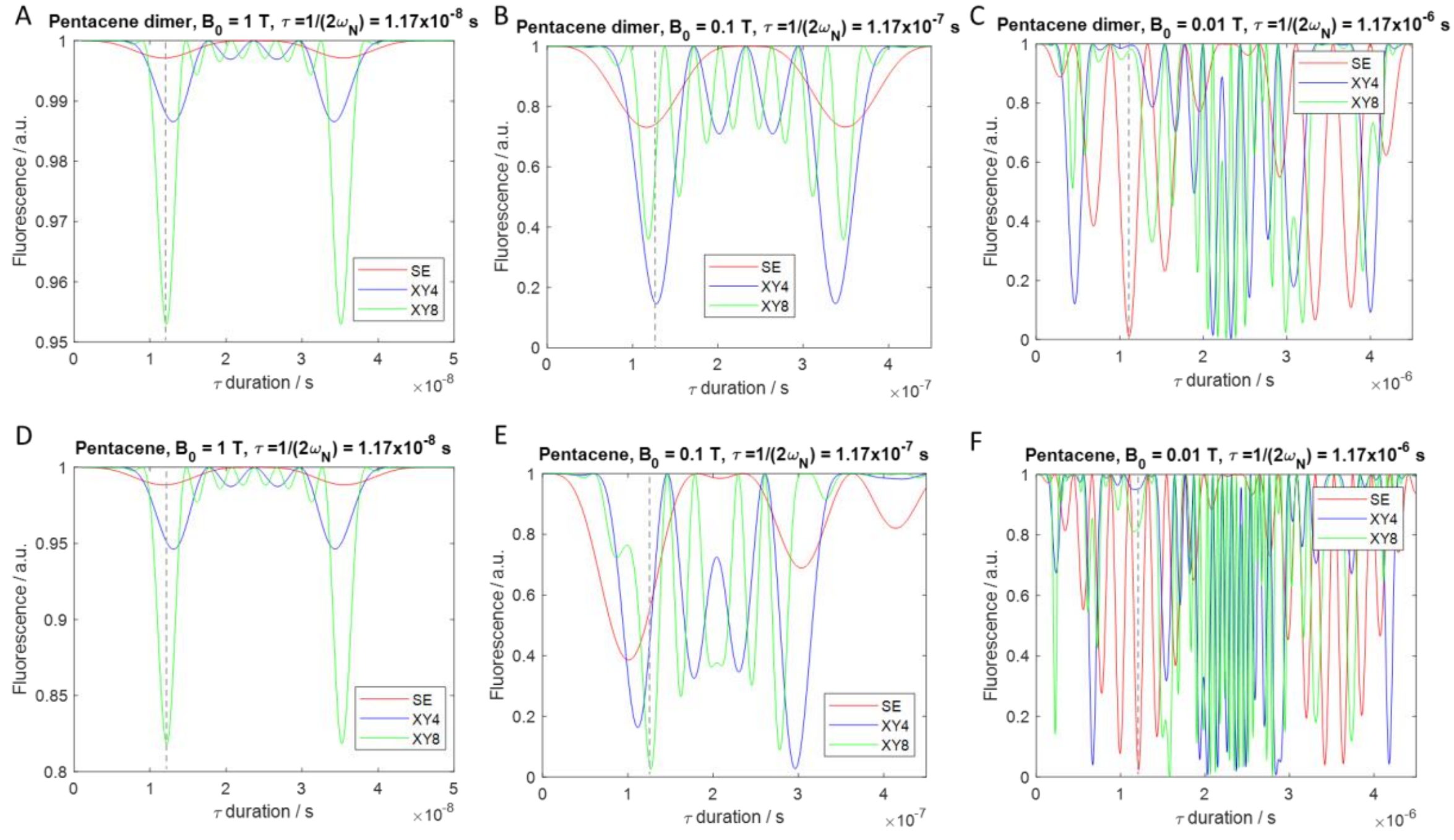

**Figure 4:** Evolution of the fluorescence as a function of the duration of the τ delay, using the XY8, XY4 and SE sequences, at $B_0$ set equal to 1 T, 0.1 T and 0.01 T, for the pentacene dimer (A-C) and monomer (D-F). The grey dashed bar indicates the condition τ = $1/2\omega_N$. Simulations were performed using the parameters in the Table 1. Relaxation is not included. For the pentacene dimer, distances between the proton and the two pentacene centers are 3 Å and 16 Å, respectively. For the pentacene monomer, the distance between the proton and the pentacene center is 3 Å.

Consistently with the literature, [3, 4, 8] the dips in the fluorescence occur at multiples of τ = $1/2\omega_N$ (τ = $3/2\omega_N$, τ = $5/2\omega_N$, etc.). However, the position of the dips and its multiplet structure are determined also by the secular and pseudo-secular components of the hyperfine interaction. In Figure 4, we observe a decrease of the fluorescence with the magnetic field $B_0$; it is more pronounced in the SE sequence and less in the XY8 sequence. We attempt to determine an analytical description of the fluorescence as a function of the τ delay and the other simulation parameters in DD sequences. We manage to obtain a short and compact expression in the case of the SE sequence for the pentacene monomer:

$$\hat{T}_0(\tau) = 1 - \frac{{A_\perp}^2 \sin\left(\frac{\tau\omega_N}{2}\right)^2 \sin\left(\frac{1}{2}\tau\sqrt{{A_\parallel}^2+{A_\perp}^2+2A_\parallel\omega_N+\omega_N^2}\right)^2}{{A_\parallel}^2+{A_\perp}^2+2A_\parallel\omega_N+\omega_N^2} \tag{4}$$

and for the pentacene dimer:

$$^5\left(\hat{T}\hat{T}\right)_0(\tau) = \frac{\left\{4\left(A_{1\parallel}+A_{2\parallel}\right)^2+3(A_{1\perp}+A_{2\perp})^2+(A_{1\perp}+A_{2\perp})^2\left[\cos\left(\frac{1}{2}\tau\sqrt{\left(A_{1\parallel}+A_{2\parallel}\right)^2+(A_{1\perp}+A_{2\perp})^2+4\omega_N\left(A_{1\parallel}+A_{2\parallel}+\omega_N\right)}\right)+2\cos(\tau\omega_N)\sin\left(\frac{1}{4}\tau\sqrt{\left(A_{1\parallel}+A_{2\parallel}\right)^2+(A_{1\perp}+A_{2\perp})^2+4\omega_N\left(A_{1\parallel}+A_{2\parallel}+\omega_N\right)}\right)^2\right]+16\omega_N\left(A_{1\parallel}+A_{2\parallel}+\omega_N\right)\right\}}{\left[4\left(A_{1\parallel}+A_{2\parallel}\right)^2+4(A_{1\perp}+A_{2\perp})^2+16\omega_N A_{1\parallel}+16\omega_N A_{2\parallel}+16\omega_N^2\right]} \tag{5}$$

The fluorescence depends on the terms that involve the nuclear spin: $\omega_N$, $A_\parallel$ and $A_\perp$, corresponding to the nuclear Larmor frequency, the secular and pseudo-secular hyperfine interaction respectively. The above equations consider a triplet pair and one single proton placed in different positions and distances away from the pentacene ring. The distribution of protons in space is shown in the section D in the SI, and the magnitude of $A_\parallel$ and $A_\perp$ depends on the proton coordinates.

The fluorescence as a function of the τ delay in the SE, XY4 and XY8 experiments is rather comparable for the pentacene and the pentacene dimer. The decay of the dip intensity with the $B_0$ is due to the presence of $\omega_N^2$ in the denominator of Eq. 4 and 5 respectively. The evolution of the fluorescence with the τ delay depends also on both $A_\parallel$ and $A_\perp$, in particular on $A_\perp$. In the case of $A_\perp$= 0 MHz, no dips are present.

The increase of the fluorescence in the sequences XY4 and XY8 is due to an increase of terms corresponding to the product of a large number of sine and cosine functions in the numerator of their analytical expressions. The pentacene monomer and the pentacene dimer show a comparable fluorescence at the lower magnetic field (0.1 T and 0.01 T), while fluorescence of slightly smaller for the pentacene dimer at 1 T. This is due to the presence of the term $16\omega_N^2$ in Eq. 5, instead of $\omega_N^2$ in Eq. 4. This explains the larger decrease of the sensitivity of the pentacene dimer with the magnetic field. The agreement between Eq. 4 and 5, and the numerical results are shown in the section C in the SI.

Chemically-bonded protons have a negligible effect on the fluorescence, as these protons have a large Fermi contact interaction $A_{\text{iso}}$ with the triplets (~ 60 MHz [35]). Simulations show that the fluorescence reduces to near zero when $A_{\text{iso}}$ approaches 60 MHz (see section E in the SI for details). For the non-bonded protons, we determine that only the protons with $A_{\perp} \neq 0$ (in the X/Z plane) contribute to the dips in the fluorescence (see Figure S3 in the SI).

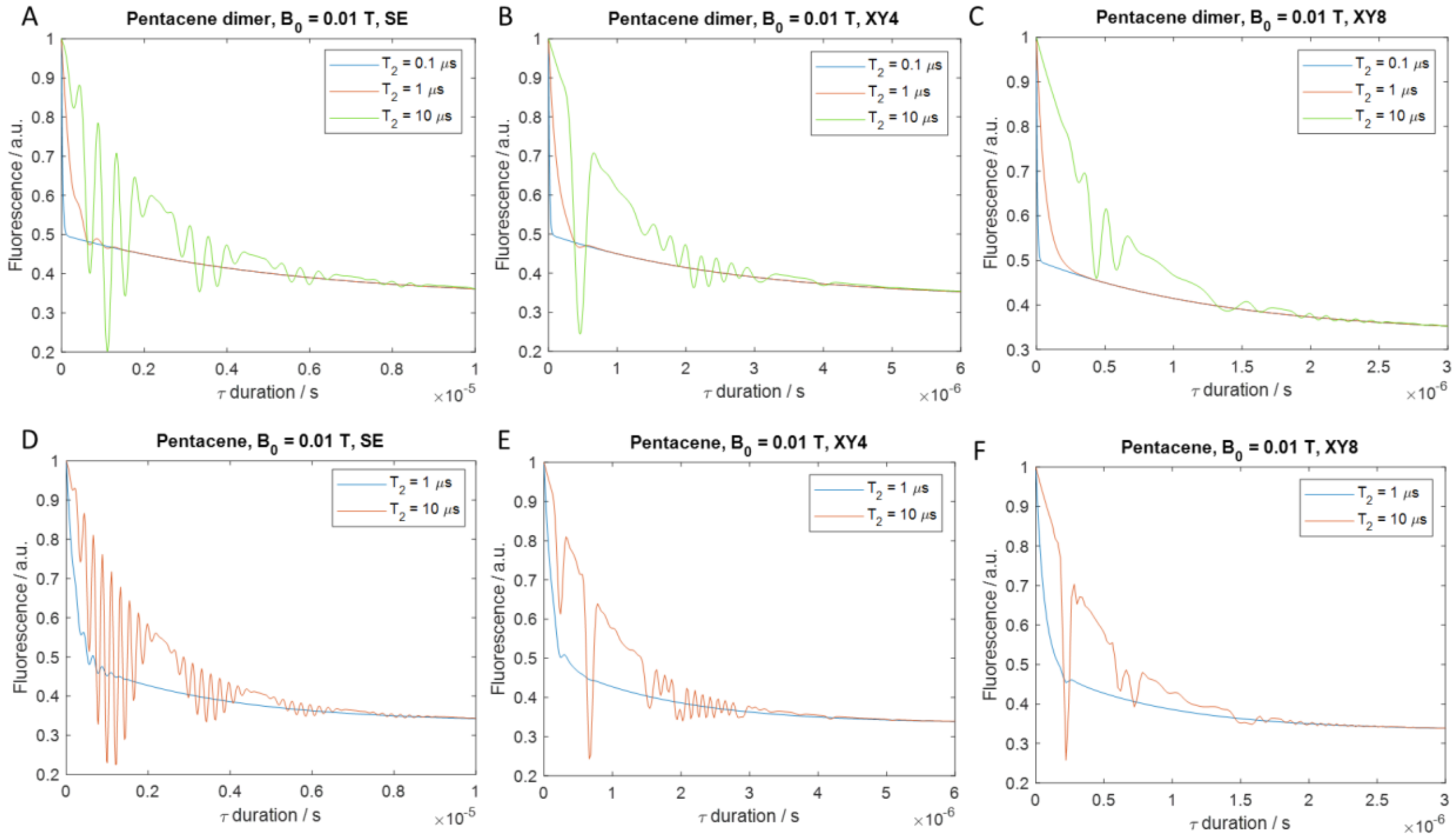

**Figure 5:** Evolution of the fluorescence as a function of the duration of the τ delay, using the XY8, XY4 and SE sequences, at $B_0$ set equal to 0.01 T, for the pentacene dimer (A-C) and monomer (D-F). Simulations were performed using the decoherence relaxation time $T_2$ given in the title and the longitudinal relaxation time $T_1$ indicated in spin systems and methodology section. The other parameters are given in the Table 1. For the pentacene dimer, distances between the proton and the two pentacene centers are 3 Å and 16 Å, respectively. For the pentacene monomer, the distance between the proton and the pentacene center is 3 Å.

We further studied the effect of the relaxation on the fluorescence under the magnetic field of 0.01 T, where we observe a stronger sensitivity (Figure 5). With relaxation, we observe a decay of the fluorescence and a decrease of the dip intensity with the increase of the duration of the τ delay. These effects are more pronounced in the XY8 sequence due to the lager number of τ delays that allow the relaxation to act for longer times. The decoherence time, acting in the X/Y plane, competes with the pseudo-secular component of the hyperfine interaction (see Eq. S10 and S24 in the SI), leading to a more pronounced fluorescence decay between 0.1 μs – 1 μs.

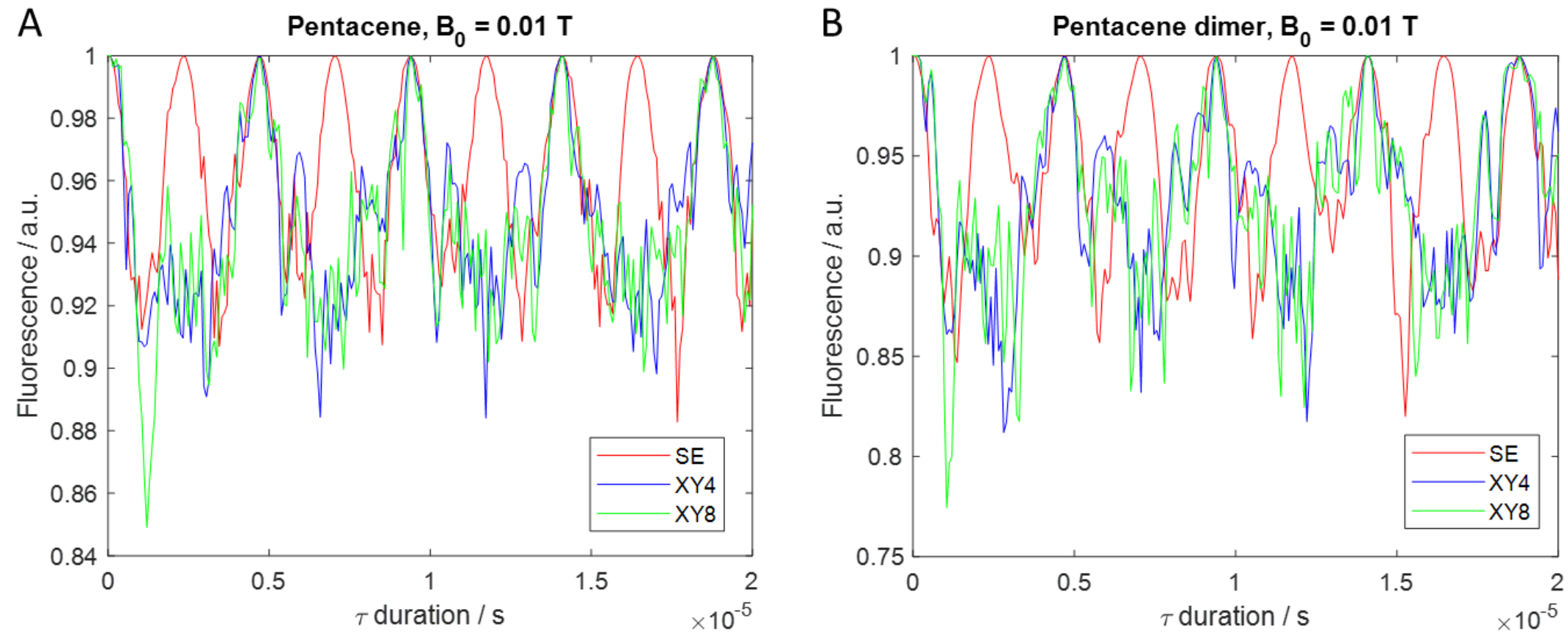

**Figure 6:** Evolution of the average fluorescence as a function of the duration of the τ delay, obtained from the pentacene monomer (A) and the pentacene dimer (B) using the XY8, XY4 and SE sequences, the $B_0$ was set equal to 0.01 T. Simulations were performed using 144 proton configurations placed within 10 Å around pentacene and the pentacene dimer. The final result corresponds to an average obtained by summing the result of 144 individual simulations performed using a triplet pair and one proton placed in different positions and distances away from one pentacene ring. The simulation parameters are given in the Table 1. Relaxation is not included. The diagram showing the configurations of the protons is shown in Figure S3B in the SI.

In the presence of multiple surrounding protons, the overall fluorescent intensity decreases but two triplets in the dimer leads to stronger fluorescence as compared with the pentacene monomer (see Figure 6). The XY8 sequence is the most sensitive to the proton signals, which shows a stronger dip at $\tau = 1/2\omega_N$, either in the absence (Figure 6) or in the presence of relaxation (Figure 7). Fourier transform (FT) of the fluorescence as a function of the time can be used to detect nuclear spins, as it manifests as a sharp peak at $\omega_N$ (see the section F in the SI). [3, 4] When the decoherence time (e.g. 10 μs) is comparable to echo time τ of pulse sequence, a clear dip can be observed; but when decoherence is fast (e.g. 1 μs), the dip disappears.

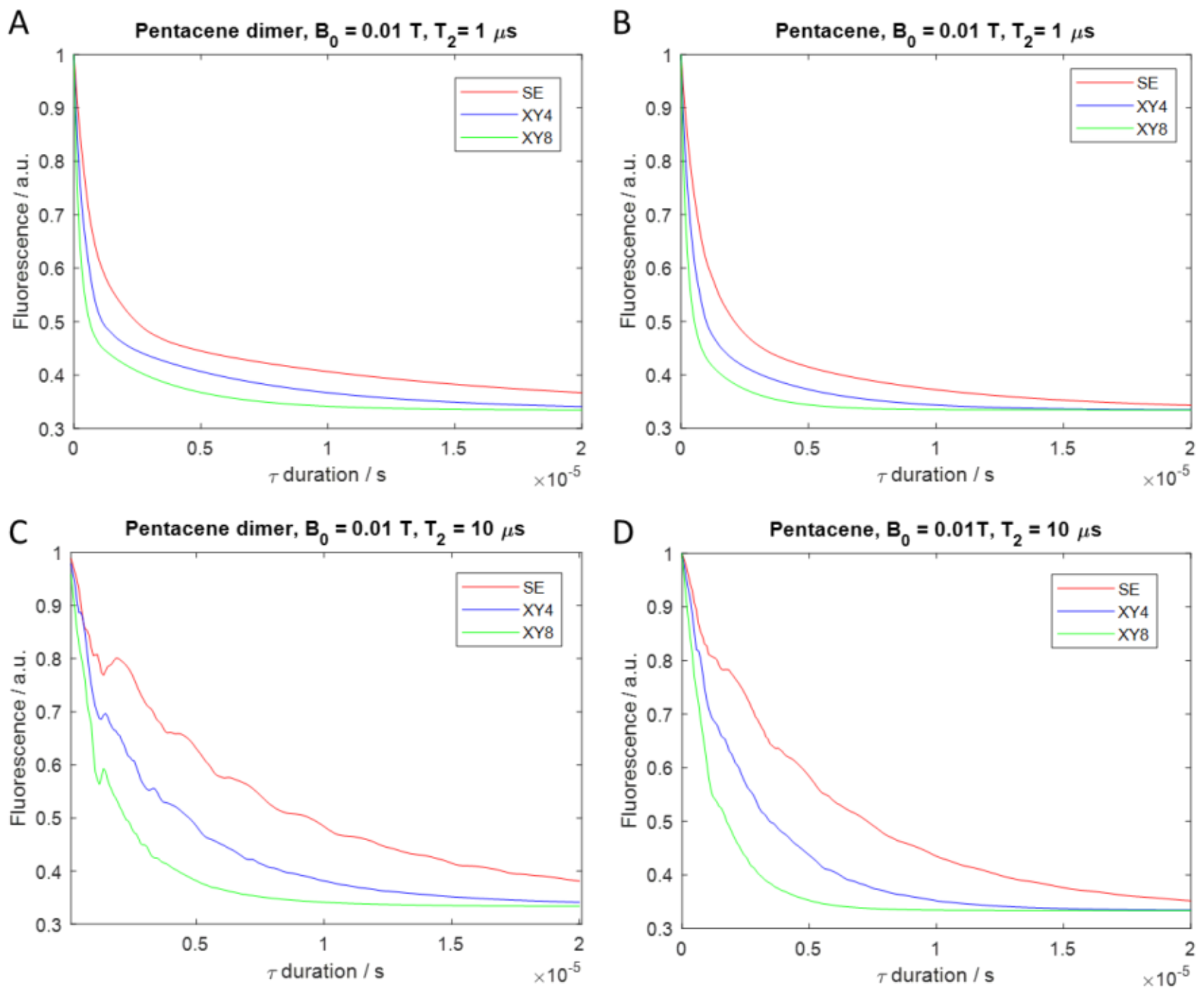

**Figure 7:** Evolution of the fluorescence as a function of the duration of the τ delay, using the XY8, XY4 and SE sequences, at $B_0$ set equal to 0.01 T, for the pentacene dimer (A and C) and monomer (B and D). Simulations were performed using two decoherence relaxation time $T_2$ given in the title and the longitudinal relaxation time $T_1$ was set to values indicated above in spin systems and methodology section. Simulations were performed using the parameters in the Table 1, the distances between the proton and the two pentacene centers are 3 Å and 16 Å, respectively.

### 3) Detection of AC field with the pentacene monomer and dimer:

The detection of AC field can be done at $B_0$>1 T where the signal is not affected by the hyperfine interaction. The AC field corresponds to a function $B_{\mathrm{AC}}\sin(2\pi\omega_{AC}t)$ applied during the DD sequences, where $B_{\mathrm{AC}}$ and $\omega_{\mathrm{AC}}$ are the power and the frequency of the AC field respectively. The AC field leads to an additional phase to the triplet spins when they precess in the X/Y plane during the τ delay. In the case of the SE sequence, the phase acquired during the first and the second $\tau$ delay in the correspond to $\theta_a$ and $\theta_b$ respectively:

$$\theta_a = B_{\mathrm{AC}} \int_0^{\tau} \sin(2\pi\omega_{AC}t)dt = \frac{B_{\mathrm{AC}}\sin(\pi\tau\omega_{AC})^2}{\pi\omega_{\mathrm{AC}}} \tag{6}$$

$$\theta_b = B_{\mathrm{AC}} \int_{\tau}^{2\tau} \sin(2\pi\omega_{\mathrm{AC}}t)\, dt = \frac{B_{\mathrm{AC}}(1+2\cos(\pi\tau\omega_{AC}))\sin(\pi\tau\omega_{AC})^2}{\pi\omega_{\mathrm{AC}}} \tag{7}$$

Ignoring the contribution of the hyperfine interaction for simplicity, the corresponding evolution of the states $\hat{T}_0$ and ${}^5(\hat{T}\hat{T})_0$ during the SE sequence can be described by the following analytical equations:

$$\hat{T}_0(\tau) = {}^5(\hat{T}\hat{T})_0(\tau) = \cos[\tfrac{1}{2}(\theta_a - \theta_b)\tau]^2 \tag{8}$$

Similarly, for the XY4 and XY8 sequences, we obtain the phases acquired during the four and eight τ delays, they are shown in the section G of the SI. The corresponding evolution of the states $\hat{T}_0$ and $^5(\hat{T}\hat{T})_0$ during the XY4 and XY8 sequences can be described by the following analytical Eqs. 9 and 10 respectively:

$$\hat{T}_0(\tau) = {}^5(\hat{T}\hat{T})_0(\tau) = \cos\,[\tfrac{1}{4}(\theta_a - 2(\theta_b - \theta_c + \theta_d) + \theta_e)\tau]^2 \qquad (9)$$

and:

$$\hat{T}_0(\tau) = {}^5(\hat{T}\hat{T})_0(\tau) = \cos[\tfrac{1}{4}(\theta_a - 2(\theta_b - \theta_c + \theta_d - \theta_e + \theta_f - \theta_g + \theta_h) + \theta_i)\tau]^2 \quad (10)$$

where $\theta_a$- $\theta_i$ correspond to the phase acquired during the four and eight τ delays in the XY4 and XY8 respectively, their analytical expressions are given in the section G in the SI. In according to Eqs. 8-10, we expect comparable fluorescence in the case pentacene and the pentacene dimer. Due to the presence of $\omega_{\mathrm{AC}}$ in the denominator of Eq. 6-7, the power of the $B_{\mathrm{AC}}$ field has to be of at least few orders of magnitude larger to have a variation of the phase. Figure 8A shows that in the absence of relaxation the XY8 is the most sensitive sequence to the AC field, dips in the fluorescence appears at multiples of τ = $1/2\omega_{\mathrm{AC}}$ (τ = $3/2\omega_{\mathrm{AC}}$, τ = $5/2\omega_{\mathrm{AC}}$, etc.), with increasing intensity with the duration of the τ delay.

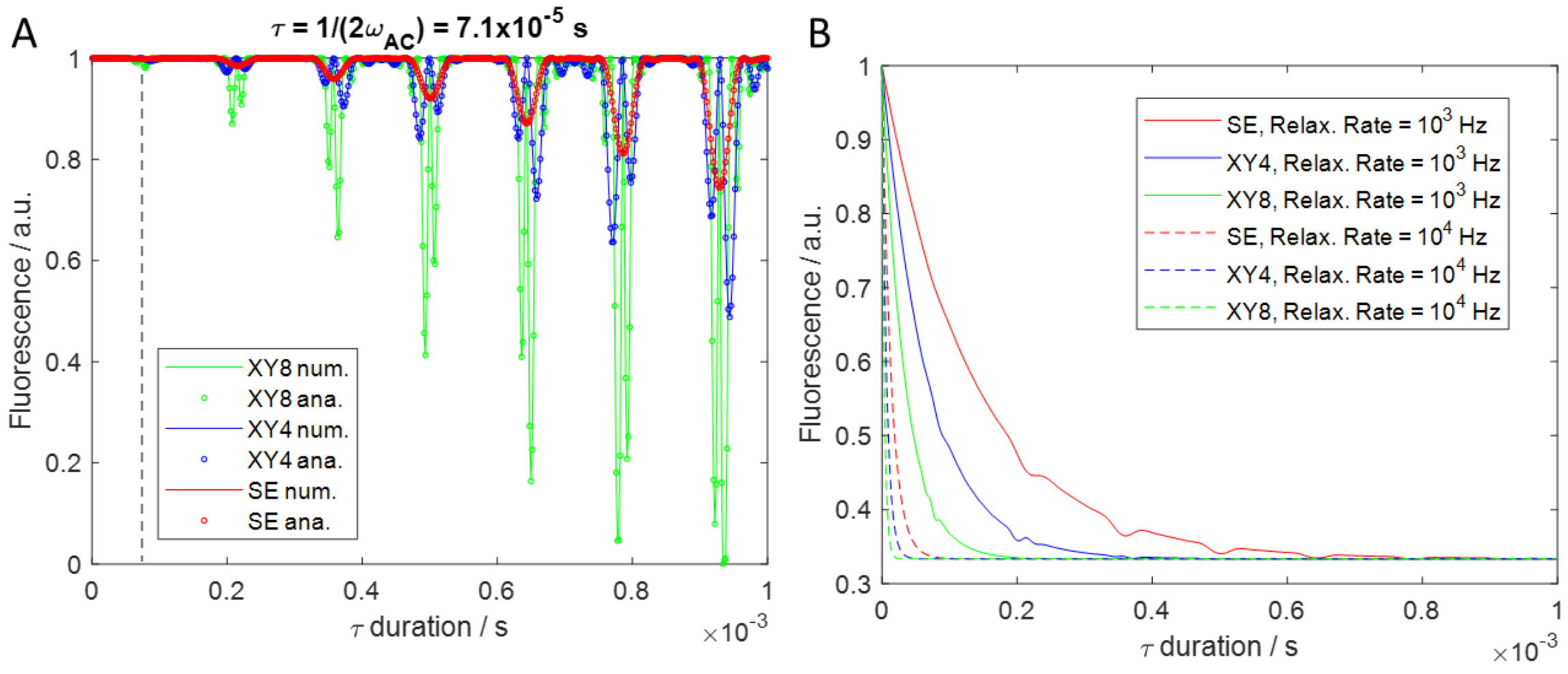

**Figure 8:** Evolution of the fluorescence as a function of the duration of the τ delay, obtained from the pentacene using the XY8, XY4 and SE sequences in the presence of an AC field, without relaxation (A) and with relaxation (B). The $B_0$ was set equal to 5 T (at which the contribution of the hyperfine is no longer observed), the $\omega_{AC}$ was set equal to $7\times10^3$ Hz corresponding to τ = $1/(2\omega_{AC})$ equal to $7.1\times10^{-5}$ s (indicated with the grey dashed line) and the $B_{AC}$ was set equal to 5 MHz. The other simulation parameters are given in the table 1, the simulation methodology is provided in the section G in the SI. Simulations were performed using the parameters in the Table 1, the distances between the proton and the pentacene center is 3 Å.

Figure 8A shows a comparison between the numerical and analytical fluorescence computed in according to Eq. 8-10. The increase of the fluorescence from SE to the XY8 is due to the increase of the terms controlling the phase of the triplet spin (i.e. $\theta_a$ and $\theta_i$) in Eq. 8-10. Their magnitude increases with the τ delay, explaining an increase of the fluorescence in Figure 8A. However, in the presence of relaxation in the order of $10^4$ Hz observed in the pentacene at room temperature, [10, 13, 36] we could not observe any dip in the fluorescence (Figure 8B). Dips should be observable for slow relaxation in the order of $10^3$ Hz when the sample is tested at cryogenic temperatures [37].

**Conclusion and outlooks:**

We analyzed the quantum sensing properties of pentacene monomers and dimers using Lindblad master equation-based spin dynamics simulations. Similar to monomers [10, 13], pentacene dimers enable coherent spin control. Single-crystal ODMR simulations revealed well-defined dips for the $(\hat{T}\hat{T})_0 \rightarrow {}^5(\hat{T}\hat{T})_{\pm1}$ and ${}^5(\hat{T}\hat{T})_{-1} \rightarrow {}^5(\hat{T}\hat{T})_{-2}$ transitions, whereas powder simulations yielded broader spectra predominantly shifted by the grid's β angle. The ODMR position and line-shape are dominated by the inter-electron exchange coupling ($J_{ex}$), which prevents the asymmetric line-shapes typically observed in monomers [10, 12, 35].

During dynamical decoupling (DD), fluorescence decays as the magnetic field increases; however, the XY8 sequence proved most sensitive to nuclear spins at fields ≤ 0.01 T. Using derived analytical expressions, we fully characterized the fluorescence behavior across SE, XY4, and XY8 sequences. While monomers and dimers exhibit comparable performance in detecting isolated nuclear spins and AC fields, the dimer's triplet entanglement provides superior sensitivity for detecting small nuclear spin ensembles, yielding significantly deeper fluorescence dips at 0.01 T.

These sensing properties depend heavily on $B_0$ and nuclear spin parameters, particularly $\omega_N$ and the pseudo-secular components of the hyperfine interaction. Notably, fluorescence is unaffected by ZFS parameters, $J_{ex}$, or inter-electron dipolar interactions. Because the coherence time ($T_2$) competes with the pseudo-secular hyperfine interactions that generate fluorescence dips, shorter $T_2$ times reduce overall sensitivity.

Based on our previous DNP studies [14], $T_2$ can be prolonged by reducing inter-triplet dipolar interactions. Future molecular designs could achieve this by synthesizing longer pentacene dimers with phenylene/acetylene linkers [38], or by utilizing monomers with more isotropic g-tensors and smaller D parameters to extend the longitudinal relaxation time ($T_1$) [39] [40]. Moving forward, we plan to investigate relaxation mechanisms, crystal orientation effects, and novel sensing protocols.

**Acknowledgments:**

This work was supported by the National Natural Science Foundation of China (22425402, 92557304 and 22275159) and Cyrus Tang Foundation (202523).

**Supplementary Information**

**Quantum Sensing with Triplet Pair States: A Theoretical Study**

Maria Grazia Concilio[1]*, Yiwen Wang[1], Siyuan Wang[1], Xueqian Kong[1]*

[1]*Institute of Translational Medicine, Shanghai Jiao Tong University, 200240 Shanghai, China*

## A. Simulation methodology

Eq. 1 in the main text can be formulated in the Liouville space in according to:

$$\frac{\partial}{\partial t}\vec{\rho} = \hat{\hat{L}}\vec{\rho} = \left(\hat{\hat{H}} + \hat{\hat{K}} + \hat{\hat{R}}\right)\vec{\rho} \tag{S1}$$

where $\hat{\hat{L}}$ is the Lindblad superoperator acting on the vectorized density matrix $\vec{\rho}$, $\hat{\hat{H}}$ is the spin Hamiltonian superoperator computed in according to:

$$\hat{\hat{H}} = -i\left(\hat{H} \otimes \mathbb{1} - \mathbb{1} \otimes \hat{H}\right) \tag{S2}$$

where $\hat{H}$ is the spin Hamiltonian operator in the Hilbert space, $\mathbb{1}$ is the unit matrix of the same size of $\hat{\hat{H}}$; $\hat{\hat{K}}$ and $\hat{\hat{R}}$ are the kinetics and the relaxation superoperator respectively, computed in according to:

$$\hat{\hat{K}} + \hat{\hat{R}} = \sum_k \left(\mathcal{L}_k^{\dagger} \otimes \mathcal{L}_k - \frac{1}{2}\mathbb{1} \otimes \left(\mathcal{L}_k^{\dagger}\mathcal{L}_k\right) - \frac{1}{2}\left(\mathcal{L}_k^{\dagger}\mathcal{L}_k\right) \otimes \mathbb{1}\right) \tag{S3}$$

where the *k* collapse operators are calculated with:

$$\mathcal{L}_k = \sqrt{R_k}\, |\psi_b\rangle\langle\psi_a| \tag{S4}$$

where $R_k$ is the rate of the transition from a state $\psi_a$ to a state $\psi_b$, corresponding to the eigenstates obtained through diagonalization of the spin Hamiltonian computed in the Hilbert space. The latter is a super-matrix composed by subspaces belonging to the states $\hat{S}_0\hat{S}_0$, $\hat{S}_1\hat{S}_0$, ${}^{\mathrm{M}}\left(\hat{T}\hat{T}\right)_{\mathrm{m_s}}$ and $\hat{T}_1 + \hat{T}_1$. In according to the scheme shown in Figure 1 in the main text, we have 12 eigenstates corresponding to the states $\hat{S}_0\hat{S}_0$, $\hat{S}_1\hat{S}_0$ $\hat{T}_1 + \hat{T}_1$, and nine ${}^{\mathrm{M}}\left(\hat{T}\hat{T}\right)_{\mathrm{m_s}}$ states. The overall SF process shown in Figure 1B in the main text, was simulated using the following equations:

$$\frac{\partial[\hat{S}_0\hat{S}_0]}{\partial t} = -k_{hv}\left[\hat{S}_0\hat{S}_0\right] + k_{flu}\left[\hat{S}_1\hat{S}_0\right] + k_{phos}\left[{}^{1}\left(\hat{T}\hat{T}\right)_0\right] + k_{rec}\left[\hat{T}_1 + \hat{T}_1\right]$$

$$\frac{\partial[\hat{S}_1\hat{S}_0]}{\partial t} = +k_{hv}\left[\hat{S}_0\hat{S}_0\right] + k_{fus}\left[{}^{1}\left(\hat{T}\hat{T}\right)_0\right] - \left(k_{flu} + k_{fis}\right)\left[\hat{S}_1\hat{S}_0\right]$$

$$\frac{\partial[{}^{1}(\hat{T}\hat{T})_0]}{\partial t} = -\left(k_{fus} + k_{phos} + x_1k_3 + x_2k_3 + x_3k_3\right)\left[{}^{1}\left(\hat{T}\hat{T}\right)_0\right] + k_{fis}\left[\hat{S}_1\hat{S}_0\right] + x_1k_{-3}\left[{}^{5}\left(\hat{T}\hat{T}\right)_0\right] + x_2k_{-3}\left[{}^{5}\left(\hat{T}\hat{T}\right)_{-1}\right] + x_3k_{-3}\left[{}^{5}\left(\hat{T}\hat{T}\right)_{-2}\right]$$

$$\frac{\partial[{}^{5}(\hat{T}\hat{T})_0]}{\partial t} = -\left(k_{dis} + x_1k_{-3} + w_{0\pm1}\right)\left[{}^{5}\left(\hat{T}\hat{T}\right)_0\right] + x_1k_3\left[{}^{1}\left(\hat{T}\hat{T}\right)_0\right] + w_{0\pm1}\left[{}^{5}\left(\hat{T}\hat{T}\right)_{\pm1}\right]$$

$$\frac{\partial[{}^{5}(\hat{T}\hat{T})_{+1}]}{\partial t} = -\left(k_{dis} + w_{0+1} + w_{-1+1} + w_{+1+2}\right)\left[{}^{5}\left(\hat{T}\hat{T}\right)_{+1}\right] + w_{0,+1}\left[{}^{5}\left(\hat{T}\hat{T}\right)_0\right] + w_{-1+1}\left[{}^{5}\left(\hat{T}\hat{T}\right)_{-1}\right] + w_{+1+2}\left[{}^{5}\left(\hat{T}\hat{T}\right)_{+2}\right]$$

$$\frac{\partial[{}^{5}(\hat{T}\hat{T})_{-1}]}{\partial t} = -\left(k_{dis} + x_2k_{-3} + w_{0-1} + w_{-1+1} + w_{-1-2}\right)\left[{}^{5}\left(\hat{T}\hat{T}\right)_{-1}\right] + w_{0,-1}\left[{}^{5}\left(\hat{T}\hat{T}\right)_0\right] + w_{-1+1}\left[{}^{5}\left(\hat{T}\hat{T}\right)_{+1}\right] + w_{-1-2}\left[{}^{5}\left(\hat{T}\hat{T}\right)_{-2}\right] + x_2k_3\left[{}^{1}\left(\hat{T}\hat{T}\right)_0\right]$$

$$\frac{\partial\left[{}^{5}\left(\hat{T}\hat{T}\right)_{+2}\right]}{\partial t} = -\left(k_{dis} + w_{+1+2}\right)\left[{}^{5}\left(\hat{T}\hat{T}\right)_{+2}\right] + w_{+1+2}\left[{}^{5}\left(\hat{T}\hat{T}\right)_{+1}\right]$$

$$\frac{\partial\left[{}^{5}\left(\hat{T}\hat{T}\right)_{-2}\right]}{\partial t} = -\left(k_{dis} + x_3k_{-3} + w_{-1-2}\right)\left[{}^{5}\left(\hat{T}\hat{T}\right)_{-2}\right] + w_{-1-2}\left[{}^{5}\left(\hat{T}\hat{T}\right)_{-1}\right] + x_3k_3\left[{}^{1}\left(\hat{T}\hat{T}\right)_0\right]$$

$$\frac{\partial[\hat{T}_1 + \hat{T}_1]}{\partial t} = -k_{rec}\left[\hat{T}_1 + \hat{T}_1\right] + k_{diss}\left({}^{5}\left(\hat{T}\hat{T}\right)_0 + {}^{5}\left(\hat{T}\hat{T}\right)_{\pm1} + {}^{5}\left(\hat{T}\hat{T}\right)_{\pm2}\right)$$

(S5)

The $\hat{\hat{K}}$ superoperator is built summing together k = 17 collapse operators corresponding to the transitions between: $\hat{S}_0\hat{S}_0 \leftrightarrow \hat{S}_1\hat{S}_0$, $\hat{S}_1\hat{S}_0 \leftrightarrow {}^1(\hat{T}\hat{T})_0$, ${}^1(\hat{T}\hat{T})_0 \rightarrow \hat{S}_0\hat{S}_0$, ${}^1(\hat{T}\hat{T})_0 \leftrightarrow {}^5(\hat{T}\hat{T})_0$, ${}^1(\hat{T}\hat{T})_0 \leftrightarrow {}^5(\hat{T}\hat{T})_{-1}$, ${}^1(\hat{T}\hat{T})_0 \leftrightarrow {}^5(\hat{T}\hat{T})_{-2}$, ${}^5(\hat{T}\hat{T})_{0,\pm1,\pm2} \rightarrow \hat{T}_1 + \hat{T}_1$ and $\hat{T}_1 + \hat{T}_1 \rightarrow \hat{S}_0\hat{S}_0$. The $\hat{\hat{R}}$ superoperator is built by summing together k = 10 collapse operators corresponding to the transitions between: ${}^5(\hat{T}\hat{T})_0 \leftrightarrow {}^5(\hat{T}\hat{T})_{\pm1}$, ${}^5(\hat{T}\hat{T})_{\pm1} \leftrightarrow {}^5(\hat{T}\hat{T})_{\pm2}$ and $(\hat{T}\hat{T})_{+1} \leftrightarrow {}^5(\hat{T}\hat{T})_{-1}$.

The laboratory frame Hamiltonian of the pentacene dimer is a very dense matrix due the presence of many off-diagonal elements arising from the anisotropic component of the Zero Field Splitting (ZFS) and inter-triplet dipolar interaction. This leads to difficulties in the diagonalization, and too complicated eigenvalue expressions to describe the ODMR contrast. Therefore, simulations were performed using a pseudo-secular Hamiltonian composed by two triplets and one proton which corresponds to:

$$\begin{aligned}
\hat{H} &= \hat{H}_{\mathrm{Z,E1}} + \hat{H}_{\mathrm{Z,E2}} + \hat{H}_{\mathrm{Z,N}} + \hat{H}_{\mathrm{ZFS1}} + \hat{H}_{\mathrm{ZFS2}} + \hat{H}_{\mathrm{d}} + \hat{H}_{\mathrm{Jex}} + \hat{H}_{\mathrm{HF1}} + \hat{H}_{\mathrm{HF2}} \\
&= \omega_E\left(\hat{E}_{Z1} + \hat{E}_{Z2}\right) + \omega_N \hat{N}_Z + D\left[\hat{E}_{Z1}^2 - \frac{1}{3}\left(\hat{E}_{Z1}^2 + \hat{E}_{X1}^2 + \hat{E}_{Y1}^2\right)\right] + \mathrm{E}\left(\hat{E}_{X1}^2 - \hat{E}_{Y1}^2\right) + \\
&\quad + D\left[\hat{E}_{Z2}^2 - \frac{1}{3}\left(\hat{E}_{Z2}^2 + \hat{E}_{X2}^2 + \hat{E}_{Y2}^2\right)\right] + \mathrm{E}\left(\hat{E}_{X2}^2 - \hat{E}_{Y2}^2\right) \\
&\quad + d\left[\hat{E}_{Z1}\cdot\hat{E}_{Z2} - \frac{1}{4}\left(\hat{E}_{+1}\cdot\hat{E}_{-2} + \hat{E}_{-1}\cdot\hat{E}_{+2}\right)\right] \\
&\quad + J_{\mathrm{ex}}\left(\hat{E}_{X1}\cdot\hat{E}_{X2} + \hat{E}_{Y1}\cdot\hat{E}_{Y2} + \hat{E}_{Z1}\cdot\hat{E}_{Z2}\right) \\
&\quad + A_{1\parallel}\hat{E}_{Z1}\cdot\hat{N}_Z + A_{1\perp}\hat{E}_{Z1}\cdot\hat{N}_X + A_{2\parallel}\hat{E}_{Z2}\cdot\hat{N}_Z + A_{2\perp}\hat{E}_{Z2}\cdot\hat{N}_X
\end{aligned} \tag{S6}$$

where $\omega_E=\omega_{E1}=\omega_{E2}$ is the isotropic component of electron Larmor frequency, $\omega_N$ is the nuclear Larmor frequency, where $\hat{H}_{\mathrm{Z,E1(2)}}$ and $\hat{H}_{\mathrm{Z,N}}$ represent the Zeeman interaction of the triplet and of the nuclear spin respectively, $\hat{H}_{\mathrm{ZFS1(2)}}$ represent the ZFS interaction of the triplet, $\hat{H}_{\mathrm{d}}$ represents the inter-triplet dipolar interaction, $\hat{H}_{\mathrm{Jex}}$ is the inter-triplet exchange interaction and $\hat{H}_{\mathrm{HF1(2)}}$ represent dipolar hyperfine interactions between the two triplets and the nucleus; $\omega_E$ and $\omega_N$ are the electron and the Larmor frequency respectively, D and E are the ZFS parameters, $d$ is the inter-triplet dipolar interaction, $J_{\mathrm{ex}}$ is the inter-triplet exchange coupling respectively, $A_{\parallel}$ and $A_{\perp}$ are the secular and pseudo-secular component of the dipolar hyperfine interaction respectively, and $\hat{E}_i$ and $\hat{N}_i$ whit $i$ = X, Y, Z, + and - , are the triplet and the nuclear spin operators.

ODMR simulations were performed considering only the electronic componet of the Hamiltonian in Eq. S6, since the hyperfine was seen not to change massively the energy levels. This was also observed in previous works. [1] We omitted the $J_{\mathrm{ex}}$ in simulations performed at magnetic fields below 1 T to keep the pseudo-secular approximation valid, since it has no effect on the photoluminescence. To simplify the calculations, we converted the spin Hamiltonian in Eq. S6 in the coupled basis using the relation:

$$\hat{H}_{\mathrm{coupled}} = U\cdot\hat{H}\cdot U^T \tag{S7}$$

where $U$ is the conversion matrix and $U^T$ is its transpose, the $U$ matrix corresponds to:

$U=$

$$
\begin{bmatrix}
0 & 0 & \frac{1}{\sqrt{3}} & 0 & -\frac{1}{\sqrt{3}} & 0 & \frac{1}{\sqrt{3}} & 0 & 0 & 0 & 0 & 0 & 0 & 0 & 0 & 0 & 0 & 0 \\
0 & -\frac{1}{\sqrt{2}} & 0 & \frac{1}{\sqrt{2}} & 0 & 0 & 0 & 0 & 0 & 0 & 0 & 0 & 0 & 0 & 0 & 0 & 0 & 0 \\
0 & 0 & -\frac{1}{\sqrt{2}} & 0 & 0 & 0 & \frac{1}{\sqrt{2}} & 0 & 0 & 0 & 0 & 0 & 0 & 0 & 0 & 0 & 0 & 0 \\
0 & 0 & 0 & 0 & 0 & \frac{1}{\sqrt{2}} & 0 & -\frac{1}{\sqrt{2}} & 0 & 0 & 0 & 0 & 0 & 0 & 0 & 0 & 0 & 0 \\
1 & 0 & 0 & 0 & 0 & 0 & 0 & 0 & 0 & 0 & 0 & 0 & 0 & 0 & 0 & 0 & 0 & 0 \\
0 & \frac{1}{\sqrt{2}} & 0 & \frac{1}{\sqrt{2}} & 0 & 0 & 0 & 0 & 0 & 0 & 0 & 0 & 0 & 0 & 0 & 0 & 0 & 0 \\
0 & 0 & \frac{1}{\sqrt{6}} & 0 & \frac{2}{\sqrt{6}} & 0 & \frac{1}{\sqrt{6}} & 0 & 0 & 0 & 0 & 0 & 0 & 0 & 0 & 0 & 0 & 0 \\
0 & 0 & 0 & 0 & 0 & \frac{1}{\sqrt{2}} & 0 & \frac{1}{\sqrt{2}} & 0 & 0 & 0 & 0 & 0 & 0 & 0 & 0 & 0 & 0 \\
0 & 0 & 0 & 0 & 0 & 0 & 0 & 0 & 1 & 0 & 0 & 0 & 0 & 0 & 0 & 0 & 0 & 0 \\
0 & 0 & 0 & 0 & 0 & 0 & 0 & 0 & 0 & 0 & 0 & \frac{1}{\sqrt{3}} & 0 & -\frac{1}{\sqrt{3}} & 0 & \frac{1}{\sqrt{3}} & 0 & 0 \\
0 & 0 & 0 & 0 & 0 & 0 & 0 & 0 & 0 & 0 & -\frac{1}{\sqrt{2}} & 0 & \frac{1}{\sqrt{2}} & 0 & 0 & 0 & 0 & 0 \\
0 & 0 & 0 & 0 & 0 & 0 & 0 & 0 & 0 & 0 & 0 & -\frac{1}{\sqrt{2}} & 0 & 0 & 0 & \frac{1}{\sqrt{2}} & 0 & 0 \\
0 & 0 & 0 & 0 & 0 & 0 & 0 & 0 & 0 & 0 & 0 & 0 & 0 & 0 & \frac{1}{\sqrt{2}} & 0 & -\frac{1}{\sqrt{2}} & 0 \\
0 & 0 & 0 & 0 & 0 & 0 & 0 & 0 & 0 & 1 & 0 & 0 & 0 & 0 & 0 & 0 & 0 & 0 \\
0 & 0 & 0 & 0 & 0 & 0 & 0 & 0 & 0 & 0 & \frac{1}{\sqrt{2}} & 0 & \frac{1}{\sqrt{2}} & 0 & 0 & 0 & 0 & 0 \\
0 & 0 & 0 & 0 & 0 & 0 & 0 & 0 & 0 & 0 & 0 & \frac{1}{\sqrt{6}} & 0 & \frac{2}{\sqrt{6}} & 0 & \frac{1}{\sqrt{6}} & 0 & 0 \\
0 & 0 & 0 & 0 & 0 & 0 & 0 & 0 & 0 & 0 & 0 & 0 & 0 & 0 & \frac{1}{\sqrt{2}} & 0 & \frac{1}{\sqrt{2}} & 0 \\
0 & 0 & 0 & 0 & 0 & 0 & 0 & 0 & 0 & 0 & 0 & 0 & 0 & 0 & 0 & 0 & 0 & 1
\end{bmatrix}
\tag{S8}
$$

The eigenstates of the spin Hamiltonian in Eq. S7, are: ${}^{1}(\hat{T}\hat{T})_{0,\alpha/\beta}$, ${}^{3}(\hat{T}\hat{T})_{0,\pm1,\alpha/\beta}$ and ${}^{5}(\hat{T}\hat{T})_{0,\pm1,\pm2,\alpha/\beta}$ states, corresponding to the nine triplet-pair states coupled to the α and β nuclear components. ODMR simulations were performed using a time-dependent Hamiltonian according to:

$$\hat{H}(t) = \hat{H} + 2\omega_1 \sin\left(2\pi\omega_{\mu w} t + \varphi\right)\left(\hat{E}_X + \hat{E}_Y + \hat{E}_Z\right) \tag{S9}$$

where $\omega_1$ is the μw power, which is set equal to 1 MHz. It gives well-defined Rabi oscillations between the transitions ${}^{5}(\hat{T}\hat{T})_{0} \rightarrow {}^{5}(\hat{T}\hat{T})_{\pm1}$ and ${}^{5}(\hat{T}\hat{T})_{-1} \rightarrow {}^{5}(\hat{T}\hat{T})_{-2}$ in Figure 3D - F respectively. $\omega_{\mu w}$ is the microwave frequency that matches the difference between the energies of two eigenstages of the Hamiltonian in Eq. S6. φ is the phase of the pulse that is set equal to zero. In simulations of dynamical decoupling (DD) experiments, we focused on the single quantum transition ${}^{5}(\hat{T}\hat{T})_{0,\alpha/\beta} \rightarrow {}^{5}(\hat{T}\hat{T})_{+1,\alpha/\beta}$, obtained using single transition operators. [2, 3] The subspace of the Hamiltonian in Eq. S6 that represents this transition is:

$$\left(\left|{}^{5}(\hat{T}\hat{T})_{+1,\alpha}\right\rangle, \left|{}^{5}(\hat{T}\hat{T})_{0,\alpha}\right\rangle, \left|{}^{5}(\hat{T}\hat{T})_{+1,\beta}\right\rangle, \left|{}^{5}(\hat{T}\hat{T})_{0,\beta}\right\rangle\right)$$

$$\begin{pmatrix} \begin{pmatrix} \frac{A_{1\parallel}}{4} + \frac{A_{2\parallel}}{4} - \mathrm{d} \\ -\frac{\mathrm{D}}{3} + \mathrm{J}_{\mathrm{ex}} + \omega_E + \frac{\omega_N}{2} \end{pmatrix} & 0 & \frac{A_{1\perp}+A_{2\perp}}{4} & 0 \\ 0 & -2\mathrm{d} - \frac{2\mathrm{D}}{3} + J_{\mathrm{ex}} + \frac{\omega_N}{2} & 0 & 0 \\ \frac{A_{1\perp}+A_{2\perp}}{4} & 0 & \begin{pmatrix} -\frac{A_{1\parallel}}{4} - \frac{A_{2\parallel}}{4} - \mathrm{d} \\ -\frac{\mathrm{D}}{3} + \mathrm{J}_{\mathrm{ex}} + \omega_E - \frac{\omega_N}{2} \end{pmatrix} & 0 \\ 0 & 0 & 0 & -2\mathrm{d} - \frac{2\mathrm{D}}{3} + \mathrm{J}_{\mathrm{ex}} - \frac{\omega_N}{2} \end{pmatrix} \tag{S10}$$

The $A_{\perp}$term mixes the α and β nuclear components of the triplet state ${}^{5}(\hat{T}\hat{T})_{+1}$.We also compared the performance of the pentacene dimer with those of a pentacene monomer which has been already widely investigated previously. [4, 5]

In the case of the pentacene monomer, an asymmetry in the ODMR line shape was observed. This is due to the inequality of a dipolar hyperfine interaction of 300 MHz (between the triplet and a nucleus at a given distance from the triplet) and the isotropic hyperfine $A_{\mathrm{iso}}$ interactions of 60 MHz (between the triplet and 14 intra-radical nuclei), which creates a broadening of the ODMR line shape. [4, 6, 7] We also evaluated the effect of the dipolar and scalar hyperfine interaction from 14 protons on the ODMR line shape of the pentacene dimer. This was achieved by adding a term $A_{\mathrm{iso}}(\hat{E}_{Z1}\hat{N}_Z)$ to the pseudo-secular Hamiltonian in Eq. S6. The subspace of the Hamiltonian that represents the transition ${}^{5}(\hat{T}\hat{T})_{0,\alpha/\beta} \rightarrow {}^{5}(\hat{T}\hat{T})_{+1,\alpha/\beta}$ becomes:

$$\left(\left|{}^{5}(\hat{T}\hat{T})_{+1,\alpha}\right\rangle, \left|{}^{5}(\hat{T}\hat{T})_{0,\alpha}\right\rangle, \left|{}^{5}(\hat{T}\hat{T})_{+1,\beta}\right\rangle, \left|{}^{5}(\hat{T}\hat{T})_{0,\beta}\right\rangle\right)$$

$$\begin{pmatrix} \omega_E + \frac{1}{12}\begin{pmatrix} 3A_{1\parallel} + 3A_{2\parallel} + 42A_{\mathrm{iso}} \\ -12d - 4D + 12\mathrm{J}_{\mathrm{ex}} \end{pmatrix} + 6\omega_N & 0 & \frac{A_{1\perp} + A_{2\perp}}{4} & 0 \\ 0 & -2\mathrm{d} - \frac{2\mathrm{D}}{3} + \mathrm{J}_{\mathrm{ex}} + \frac{\omega_N}{2} & 0 & 0 \\ \frac{A_{1\perp} + A_{2\perp}}{4} & 0 & \omega_E + \frac{1}{12}\begin{pmatrix} -3A_{1\parallel} - 3A_{2\parallel} - 42A_{\mathrm{iso}} \\ -12d - 4D + 12\mathrm{J}_{\mathrm{ex}} \end{pmatrix} - 6\omega_N & 0 \\ 0 & 0 & 0 & -2\mathrm{d} - \frac{2\mathrm{D}}{3} + \mathrm{J}_{\mathrm{ex}} - \frac{\omega_N}{2} \end{pmatrix} \tag{S11}$$

The eigenvalues of the matrix S11 related to the states ${}^{5}(\hat{T}\hat{T})_{+1,\alpha}$ and ${}^{5}(\hat{T}\hat{T})_{+1,\beta}$ equal to: $-\mathrm{d} - \frac{\mathrm{D}}{3} + J_{\mathrm{ex}} + \omega_E \mp \frac{1}{4}\sqrt{(A_{1\parallel} + A_{2\parallel} + 14\mathrm{A}_{\mathrm{iso}})^2 + (A_{1\perp} + A_{2\perp})^2 + 4\omega_N(A_{1\parallel} + A_{2\parallel} + 14\mathrm{A}_{\mathrm{iso}} + \omega_N)}$.

The dominant inter-triplet exchange coupling $J_{\mathrm{ex}}$ equals to 15 GHz. [8] $A_{\mathrm{iso}}$ is set to 60 MHz, [6] and $A_{1\perp(\parallel)}$ and $A_{2\perp(\parallel)}$ range between 0.0038 MHz and 2.2 MHz, calculated considering a proton placed 3 Å and 16 Å away from the first and the second triplet respectively. With these numbers, we determined that the frequency of the eigenvalues of the states ${}^{5}(\hat{T}\hat{T})_{+1,\alpha}$ and ${}^{5}(\hat{T}\hat{T})_{+1,\beta}$ changed from 14.44 GHz and 14.81 GHz to 14.60 GHz and 14.64 GHz without the presence of the isotropic hyperfine. Therefore, we determined that a negligible variation of the energy levels upon the presence of the hyperfine interaction. Therefore, we do not expect to observe an asymmetry in the ODMR line shape of the pentacene dimer.

The energies of the eigenstates, obtained through diagonalization of the Hamiltonian in Eq. S6 (where we considered only the electronic component), correspond to:

$$E_{^1(\hat{T}\hat{T})_0} = \frac{1}{6}(-6\mathrm{d} - 2\mathrm{D} - 3J_{\mathrm{ex}} - 3\sqrt{4(3\mathrm{d}^2 - 2\mathrm{d} \times \mathrm{D} + \mathrm{D}^2) - 4(3\mathrm{d} + \mathrm{D})J_{\mathrm{ex}} + 9J_{\mathrm{ex}}^2} \tag{S12}$$

$$E_{^3(\hat{T}\hat{T})_{+1}} = \mathrm{d} - \frac{\mathrm{D}}{3} - J_{\mathrm{ex}} + \omega_E \tag{S13}$$

$$E_{^3(\hat{T}\hat{T})_0} = -2\mathrm{d} + \frac{2\mathrm{D}}{3} - J_{\mathrm{ex}} \tag{S14}$$

$$E_{^3(\hat{T}\hat{T})_{-1}} = \mathrm{d} - \frac{\mathrm{D}}{3} - J_{\mathrm{ex}} - \omega_E \tag{S15}$$

$$E_{^5(\hat{T}\hat{T})_{+2}} = 2\mathrm{d} + \frac{2\mathrm{D}}{3} + J_{\mathrm{ex}} + 2\omega_E \tag{S16}$$

$$E_{^5(\hat{T}\hat{T})_{+1}} = -\mathrm{d} - \frac{\mathrm{D}}{3} + J_{\mathrm{ex}} + \omega_E \tag{S17}$$

$$E_{^5(\hat{T}\hat{T})_0} = \frac{1}{6}(-6\mathrm{d} - 2\mathrm{D} - 3J_{\mathrm{ex}} + 3\sqrt{4(3\mathrm{d}^2 - 2\mathrm{d} \times \mathrm{D} + \mathrm{D}^2) - 4(3\mathrm{d} + \mathrm{D})J_{\mathrm{ex}} + 9J_{\mathrm{ex}}^2}) \tag{S18}$$

$$E_{^5(\hat{T}\hat{T})_{-1}} = -\mathrm{d} - \frac{\mathrm{D}}{3} + J_{\mathrm{ex}} - \omega_E \tag{S19}$$

$$E_{^5(\hat{T}\hat{T})_{-2}} = 2\mathrm{d} + \frac{2\mathrm{D}}{3} + J_{\mathrm{ex}} - 2\omega_E \tag{S20}$$

The allowed transition frequencies corresponding to $^3(\hat{T}\hat{T})_0 \rightarrow {}^3(\hat{T}\hat{T})_{\pm1}$ and $^5(\hat{T}\hat{T})_{\pm1} \rightarrow {}^5(\hat{T}\hat{T})_{\pm2}$ are:

$$\Delta E_{\mathbf{3\pm10}} = -3\mathrm{d} + \mathrm{D} \mp \omega_E \text{ for } (\hat{T}\hat{T})_0 \rightarrow {}^3(\hat{T}\hat{T})_{\pm1} \tag{S21}$$

$$\Delta E_{\mathbf{5\pm2\pm1}} = 3\mathrm{d} + \mathrm{D} \pm \omega_E \text{ for } {}^5(\hat{T}\hat{T})_{\pm1} \rightarrow {}^5(\hat{T}\hat{T})_{\pm2} \tag{S22}$$

The spin Hamiltonian of a spin system is composed by one triplet and one proton corresponds to:

$$\begin{aligned} \hat{H} &= \hat{H}_{\mathrm{Z,E}} + \hat{H}_{\mathrm{Z,N}} + \hat{H}_{\mathrm{ZFS}} + \hat{H}_{\mathrm{HF}} \\ &= \omega_E \hat{E}_Z + \omega_N \hat{N}_Z + D\left[\hat{E}_Z^2 - \frac{1}{3}\left(\hat{E}_Z^2 + \hat{E}_X^2 + \hat{E}_Y^2\right)\right] + \mathrm{E}\left(\hat{E}_X^2 - \hat{E}_Y^2\right) + A_{\parallel}\hat{E}_Z \cdot \hat{N}_Z + A_{\perp}\hat{E}_Z \cdot \hat{N}_X \end{aligned} \tag{S23}$$

In case of the pentacene monomer, we consider the single quantum transition $\hat{T}_{0,\alpha/\beta} \rightarrow \hat{T}_{+1,\alpha/\beta}$. The subspace of this transition in the spin Hamiltonian, corresponds to:

$$\left(|\hat{T}_{+1,\alpha}\rangle, |\hat{T}_{+1,\beta}\rangle, |\hat{T}_{0,\alpha}\rangle, |\hat{T}_{0,\beta}\rangle\right)$$

$$\begin{pmatrix} \frac{A_{\parallel}}{2} + \frac{\mathrm{D}}{3} + \omega_E + \frac{\omega_N}{2} & \frac{A_{\perp}}{2} & 0 & 0 \\ \frac{A_{\perp}}{2} & -\frac{A_{\parallel}}{2} + \frac{\mathrm{D}}{3} + \omega_E - \frac{\omega_N}{2} & 0 & 0 \\ 0 & 0 & -\frac{2\mathrm{D}}{3} + \frac{\omega_N}{2} & 0 \\ 0 & 0 & 0 & -\frac{2\mathrm{D}}{3} - \frac{\omega_N}{2} \end{pmatrix} \tag{S24}$$

The $A_{\perp}$ term mixes the α and β nuclear components of the triplet state $\hat{T}_{+1}$. The time evolution of the vectorized density matrix $\vec{\rho}(t)$, is given by:

$$\vec{\rho}(t) = e^{\left(\hat{\hat{H}}+\hat{\hat{R}}+\hat{\hat{R}}\right)}\vec{\rho}(t=0) \quad \text{(S25)}$$

The expectation value corresponding to the evolution of a spin state in the time $\hat{O}(t)$ was computed using the Hadamard matrix product: $\hat{O}(t)$=Tr$\left[\left[\hat{O}\vec{\rho}(t)\right]\right]$, where $\hat{O}$ is the observable operator.

### B. Evolution of the transitions frequency as function of the orientation of the system toward the magnetic field.

To interpret the ODMR powder spectrum shown in Figure 3B in the main text, we computed the evolution of the transitions ${}^5(\hat{T}\hat{T})_0 \rightarrow {}^5(\hat{T}\hat{T})_{\pm1}$ and ${}^5(\hat{T}\hat{T})_{\pm1} \rightarrow {}^5(\hat{T}\hat{T})_{\pm2}$ as function of the Euler angles α, β and γ. The angles specify the orientation of the single crystal relative to the laboratory reference frame (Figure S1).

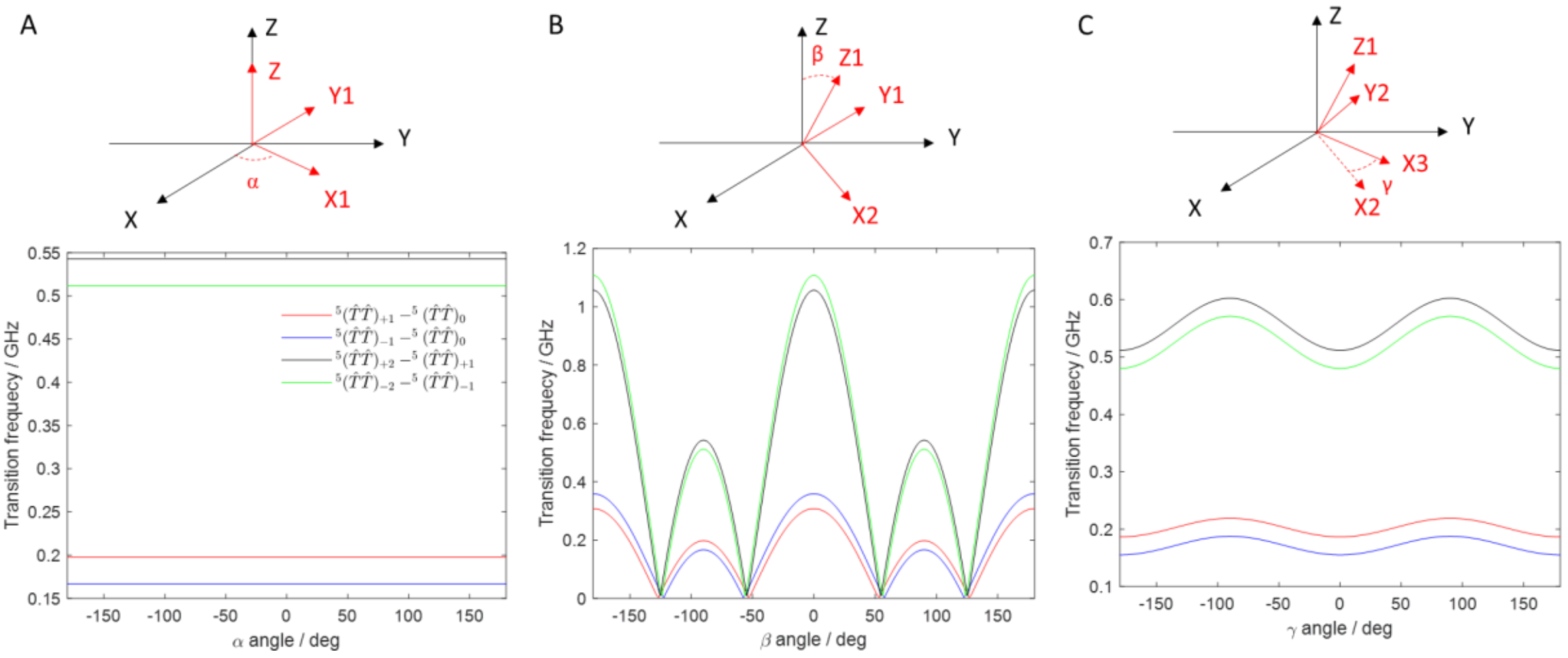

**Figure S1:** Evolution of the transitions ${}^5(\hat{T}\hat{T})_0 \rightarrow {}^5(\hat{T}\hat{T})_{\pm1}$ and ${}^5(\hat{T}\hat{T})_{\pm1} \rightarrow {}^5(\hat{T}\hat{T})_{\pm2}$ as function of the α (A), β (B) and γ (C) defined in the top panels.

We observed no variation of the transition frequency with the α angle, a slight variation with the γ angle, and a large variation of the β angle. The change of β angle leads to the overlap of the transition ${}^5(\hat{T}\hat{T})_0 \rightarrow {}^5(\hat{T}\hat{T})_{\pm1}$ and ${}^5(\hat{T}\hat{T})_{\pm1} \rightarrow {}^5(\hat{T}\hat{T})_{\pm2}$ between ±130° and ±120°, and ±60° and ±50°. This explains the shift of the transition frequencies in the Figure 3B in the main text.

## C. Agreement between the numerical and analytical result

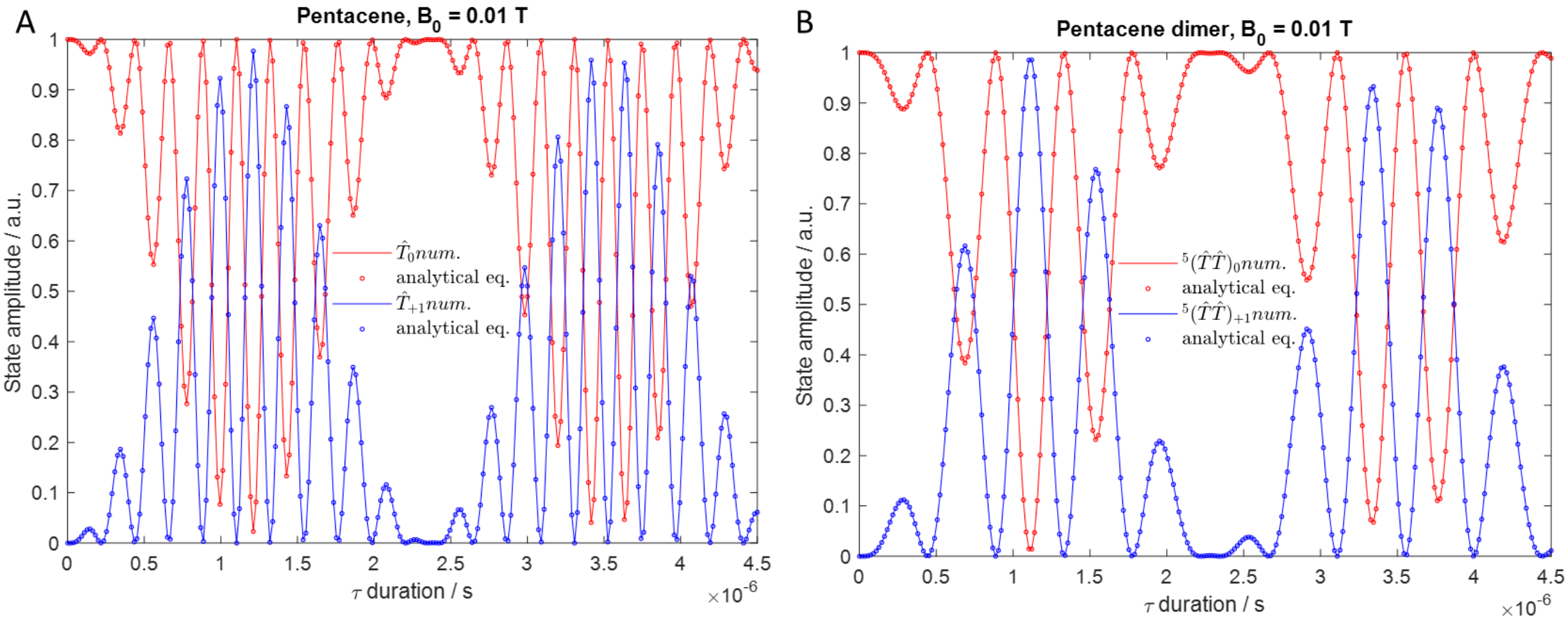

**Figure S2:** Evolution of the states $\hat{T}_{0,+1}$ and ${}^5(\hat{T}\hat{T})_{0,+1}$with the τ delay in SE sequence at $B_0$ set equal to 0.01 T for the pentacene (A) and the pentacene dimer (B). Numerical results were compared with the analytical equations (Eq. 4 for $\hat{T}_0(\tau)$ and Eq. 5 for ${}^5(\hat{T}\hat{T})_0(\tau)$ in the main text). The analytical equation of the states $\hat{T}_{+1}$ and ${}^5(\hat{T}\hat{T})_{+1}$ are obtained from $\hat{T}_{+1}(\tau) = 1 - \hat{T}_0(\tau)$ and ${}^5(\hat{T}\hat{T})_{+1}(\tau)$ = 1 - ${}^5(\hat{T}\hat{T})_0(\tau)$. Simulations were performed using the parameters in the Table 1, relaxation was not included. For the pentacene dimer, distances between the proton and the two pentacene centers are 3 Å and 16 Å, respectively. For the pentacene monomer, the distance between the proton and the pentacene center is 3 Å.

## D. Evolution of the secular and pseudo-secular components of the hyperfine interaction as a function of the proton coordinates

Figure S3A shows $A_{\parallel} \neq 0$ and $A_{\perp} \neq 0$ in the X/Z plane, while $A_{\perp} \neq 0$ and $A_{\perp} = 0$ in the X/Y plane. Therefore, only the protons present in the X/Z plane contribute to the dips in the fluorescence shown in the main text. Figure 3B shows the arrangement of the protons used in the simulations shown in Figures 6 and 7.

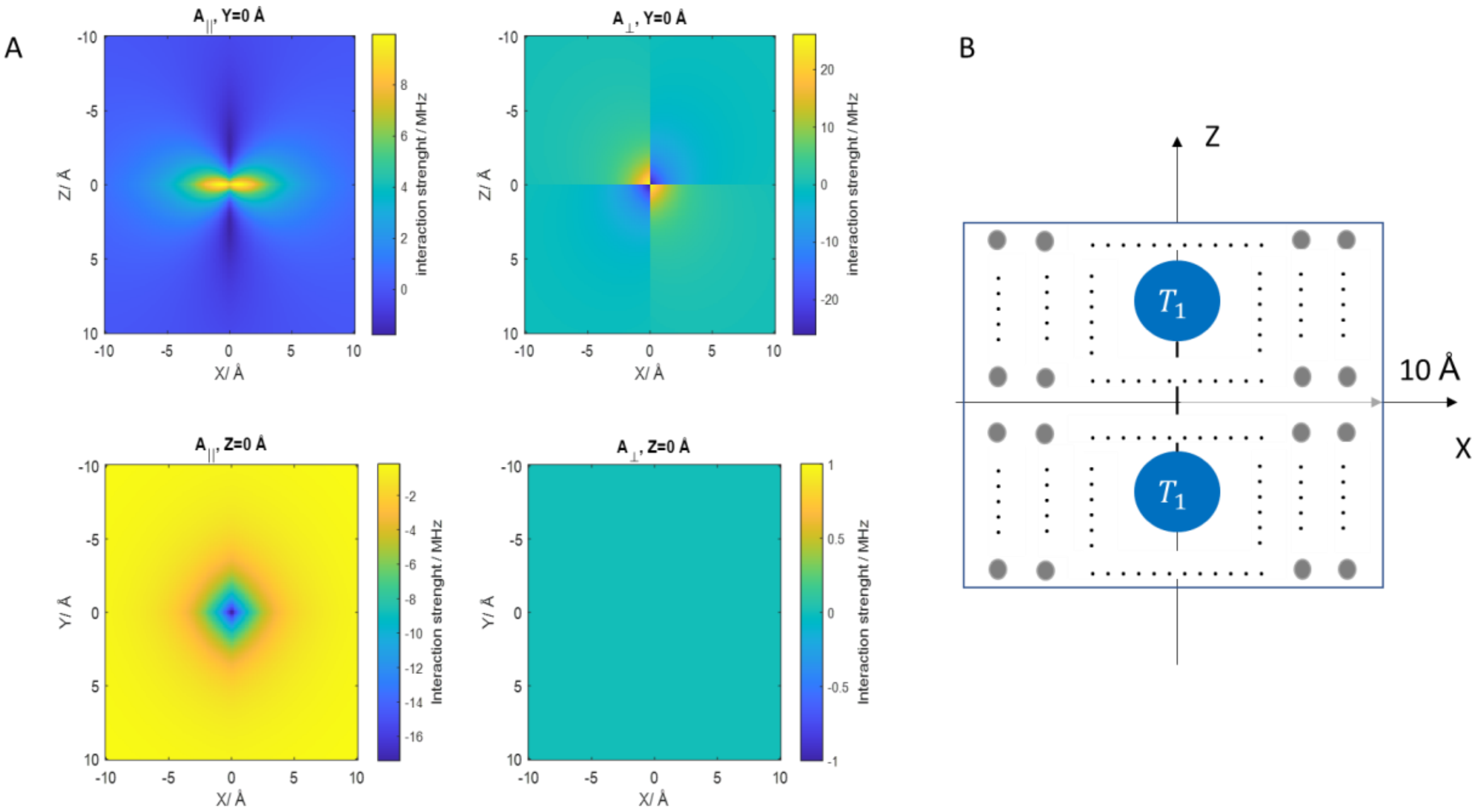

**Figure S3:** A) Magnitude of $A_{\parallel}$ and $A_{\perp}$ as a function of the coordinates of the nuclear spin placed within 10 Å from one single triplet. B) Diagram showing the position of the protons in the X/Z plane, surrounding the pentacene dimer, placed symmetrically within 10 Å the center of the box; $T_1$ represents the triplets and the grey balls represent the protons.

### E. Effect of the Fermi contact coupling on the fluorescence

In order to verify the contribution of the bonded protons on the fluorescence, we performed a set of simulations that show the evolution of the fluorescence as a function of the Fermi contact coupling $A_{iso}$ (or scalar hyperfine coupling), see Figure S4.

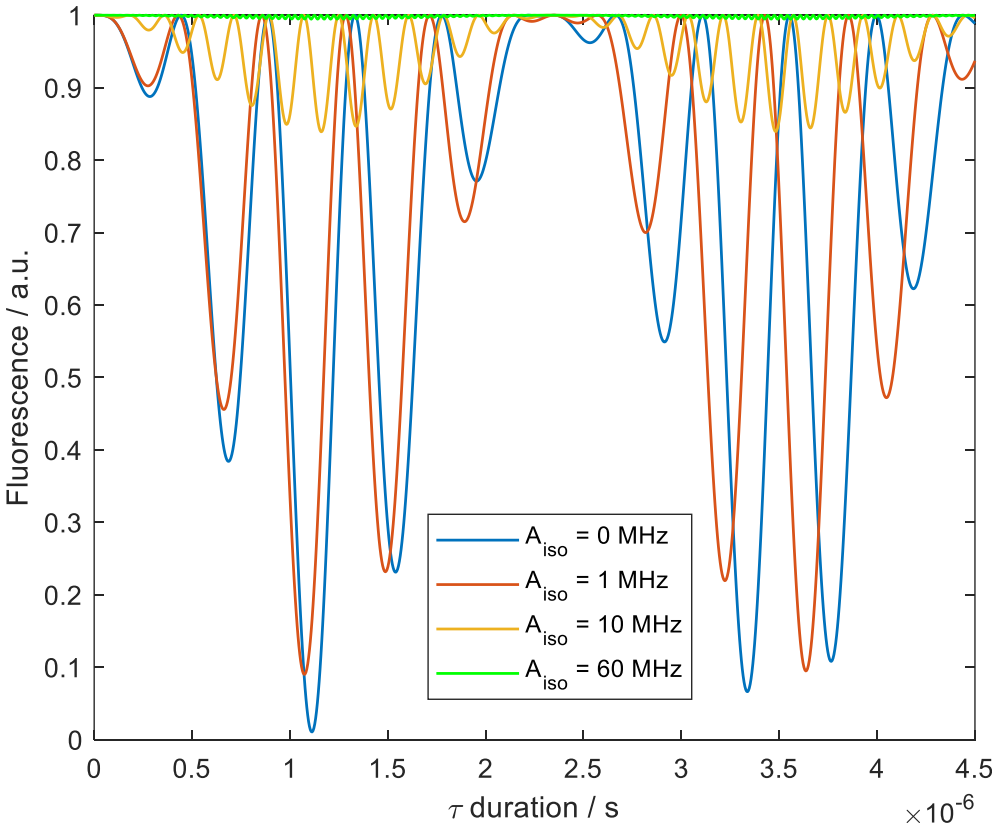

**Figure S4:** Evolution of the fluorescence as a function of the duration of the τ delay and the $A_{iso}$, using the SE sequence, at $B_0$ set equal to 0.01 T, for the pentacene dimer. Simulations were performed using the parameters in the Table 1, relaxation was not included. In the simulation, the distances between the proton and the two pentacene centers are 3 Å and 16 Å, respectively.

We observed a decrease of the fluorescence with the increase of the $A_{\text{iso}}$. The fluorescence approaches zero when $A_{\text{iso}}$ equals to 60 MHz, that is the value in pentacene. [6] Therefore, we do not expect the chemically-bonded protons to have an effect on the fluorescence.

### F. The Fourier transform (FT) of the fluorescence as a function of the time

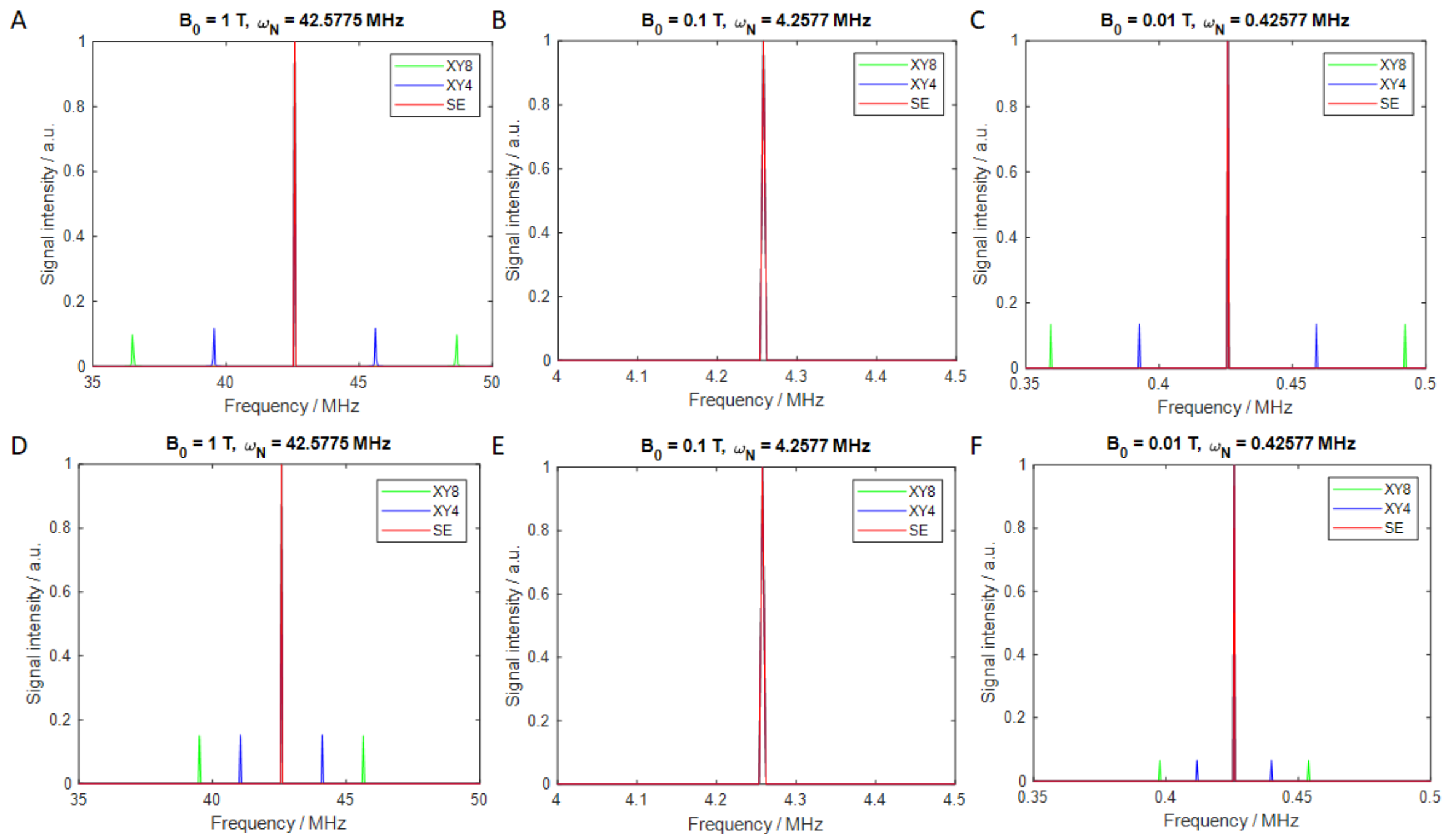

**Figure S5:** Normalized FT signal of the fluorescence in the time obtained from the XY8, XY4 and SE sequences at $B_0$ equal to 1 T, 0.1 T and 0.01 T, for the pentacene (A-C) and the pentacene dimer (D-F). Simulations were performed using a single $\tau = 1/2\omega_N$, the other simulation parameters are given in the Table 1. For the pentacene dimer, distances between the proton and the two pentacene centers are 3 Å and 16 Å, respectively. For the pentacene monomer, the distance between the proton and the pentacene center is 3 Å.

### G. Phase acquired by the triplet spin during the XY4 and XY8 sequences:

Phase acquired by the triplet spin during the XY4 and XY8 sequences can be computed in according to:

$$\theta_a = B_{AC}\int_0^{\tau/2}\sin(2\pi\omega_{AC}t)dt = \frac{B_{AC}\sin(\pi\tau\omega_{AC})^2}{\pi\omega_{AC}} \tag{S26}$$

$$\theta_b = B_{\mathrm{AC}}\int_{\tau/2}^{3\tau/2}\sin(2\pi\omega_{\mathrm{AC}}t)dt = \frac{B_{AC}\sin(\pi\tau\omega_{AC})\sin(2\pi\tau\omega_{AC})}{\pi\omega_{AC}} \tag{S27}$$

$$\theta_c = B_{\mathrm{AC}}\int_{3\tau/2}^{5\tau/2}\sin(2\pi\omega_{\mathrm{AC}}t)dt = \frac{B_{AC}[\cos(3\pi\tau\omega_{AC})-\cos(5\pi\tau\omega_{AC})]}{2\pi\omega_{AC}} \tag{S28}$$

$$\theta_d = B_{\mathrm{AC}}\int_{5\tau/2}^{7\tau/2}\sin(2\pi\omega_{\mathrm{AC}}t)dt = \frac{B_{AC}[\cos(5\pi\tau\omega_{AC})-\cos(7\pi\tau\omega_{AC})]}{2\pi\omega_{AC}} \tag{S29}$$

$$\theta_e = B_{\mathrm{AC}}\int_{7\tau/2}^{4\tau}\sin(2\pi\omega_{\mathrm{AC}}t)dt = \frac{B_{AC}[\cos(7\pi\tau\omega_{AC})-\cos(8\pi\tau\omega_{AC})]}{2\pi\omega_{AC}} \tag{S30}$$

for XY4, and for XY8:

$$\theta_a = B_{\mathrm{AC}}\int_0^{\tau/2}\sin(2\pi\omega_{\mathrm{AC}}t)dt = \frac{B_{AC}\sin(\pi\tau\omega_{AC})^2}{\pi\omega_{AC}} \tag{S31}$$

$$\theta_b = B_{\mathrm{AC}}\int_{\tau/2}^{3\tau/2}\sin(2\pi\omega_{\mathrm{AC}}t)dt = \frac{B_{AC}\sin(\pi\tau\omega_{AC})\sin(2\pi\tau\omega_{AC})}{\pi\omega_{AC}} \tag{S32}$$

$$\theta_c = B_{\mathrm{AC}}\int_{3\tau/2}^{5\tau/2}\sin(2\pi\omega_{\mathrm{AC}}t)dt = \frac{B_{AC}[\cos(3\pi\tau\omega_{AC})-\cos(5\pi\tau\omega_{AC})]}{2\pi\omega_{AC}} \tag{S33}$$

$$\theta_d = B_{\mathrm{AC}}\int_{5\tau/2}^{7\tau/2}\sin(2\pi\omega_{\mathrm{AC}}t)dt = \frac{B_{AC}[\cos(5\pi\tau\omega_{AC})-\cos(7\pi\tau\omega_{AC})]}{2\pi\omega_{AC}} \tag{S34}$$

$$\theta_e = B_{\mathrm{AC}}\int_{7\tau/2}^{9\tau/2}\sin(2\pi\omega_{\mathrm{AC}}t)dt = \frac{B_{AC}[\cos(7\pi\tau\omega_{AC})-\cos(9\pi\tau\omega_{AC})]}{2\pi\omega_{AC}} \tag{S35}$$

$$\theta_f = B_{\mathrm{AC}}\int_{9\tau/2}^{11\tau/2}\sin(2\pi\omega_{\mathrm{AC}}t)dt = \frac{B_{AC}[\cos(9\pi\tau\omega_{AC})-\cos(11\pi\tau\omega_{AC})]}{2\pi\omega_{AC}} \tag{S36}$$

$$\theta_g = B_{\mathrm{AC}}\int_{11\tau/2}^{13\tau/2}\sin(2\pi\omega_{\mathrm{AC}}t)dt = \frac{B_{AC}[\cos(11\pi\tau\omega_{AC})-\cos(13\pi\tau\omega_{AC})]}{2\pi\omega_{AC}} \tag{S37}$$

$$\theta_h = B_{\mathrm{AC}}\int_{13\tau/2}^{15\tau/2}\sin(2\pi\omega_{\mathrm{AC}}t)dt = \frac{B_{AC}[\cos(13\pi\tau\omega_{AC})-\cos(15\pi\tau\omega_{AC})]}{2\pi\omega_{AC}} \tag{S38}$$

$$\theta_i = B_{\mathrm{AC}}\int_{15\tau/2}^{8\tau}\sin(2\pi\omega_{\mathrm{AC}}t)dt = \frac{B_{AC}[\cos(15\pi\tau\omega_{AC})-\cos(16\pi\tau\omega_{AC})]}{2\pi\omega_{AC}} \tag{S39}$$

Simulations shown in Figure 8 were performed by calculating the phase acquired by the triplets using Eq. 6 and 7 in the main text and S26-S39 in the supporting information. The spin Hamiltonian $\hat{H}_i$ was then computed in according to $\hat{H}_i = \hat{H} + \theta_i\hat{E}_Z$, where $\hat{H}$ is the spin Hamiltonian containing the interactions in the system, $\theta_i$ with $i$ = *a-i* is the phase acquired by the triplets, and $\hat{E}_Z$ is the triplet operator. The τ delay propagators was then computed in according to $\hat{P}_i = e^{-i\tau\hat{H}_i}$. The final state vector $\vec{\rho}(t)$ can be obtained from $\vec{\rho}(t) = \hat{U}\vec{\rho}(t=0)$, where $\hat{U}$ is the overall propagator operator obtained multiplying the pulse and τ delay propagators. The expectation value can be computed using the Hadamard matrix product.